\documentclass[11pt]{article}
\usepackage[T1]{fontenc}
\usepackage{microtype}
\usepackage{amsmath}
\usepackage{amsfonts}
\usepackage{amssymb}
\usepackage{amsthm}
\usepackage{graphicx}
\usepackage[margin=1.00in]{geometry}
\graphicspath{{img/}}
\usepackage{booktabs}
\usepackage{tabularx}
\usepackage{hyperref}
\usepackage{breakcites}
\usepackage{placeins}
\usepackage{multicol}
\usepackage{bbm}
\usepackage{multirow}
\usepackage{subfig}
\usepackage{tikz}
\usepackage{authblk}
\usepackage{natbib}

\usepackage{xcolor}
\usepackage{colortbl}
\definecolor{modelgray}{gray}{0.93}

\tikzstyle{arrow} = [thick,->,>=stealth]
\usetikzlibrary{trees,shapes,decorations, arrows}
\usetikzlibrary{shapes,positioning,fit}

\usepackage[framemethod=TikZ]{mdframed}
\usepackage{fvextra}  

\DefineVerbatimEnvironment{prompt}{Verbatim}{
  breaklines=true,                 
  breaksymbolleft={},              
  breakanywhere=true,              
  obeytabs=true,
  codes={\catcode`\$=12 \catcode`\_=12 \catcode`\^=12} 
}

\surroundwithmdframed[
  backgroundcolor=gray!10,
  linecolor=gray!50,
  linewidth=0.5pt,
  roundcorner=2pt,
  innertopmargin=4pt,
  innerbottommargin=4pt,
  innerleftmargin=6pt,
  innerrightmargin=6pt,
  skipabove=8pt,
  skipbelow=8pt
]{prompt}

\title{Semantic insurance pricing with large language models}

\author{Christopher Blier-Wong}
\author{Derek Kusmenko}
\affil{Department of Statistical Sciences, University of Toronto, Canada}

\date{\today}

\begin{document}

\maketitle

\begin{abstract}

Classical actuarial pricing models, such as the generalized linear model, are valued for transparency and ease of governance, but they use interactions among risk factors only when these are supplied through explicit feature engineering. We study whether embeddings from a pre-trained large language model, computed from a natural-language description of each policyholder, can replace hand-crafted features as inputs to a standard actuarial pricing model, taking Poisson claim-frequency regression as the main example. The language model is used only to construct deterministic embedding covariates; pricing is performed by a standard generalized linear model. Using French motor third-party liability data, the embedding-based model outperforms the generalized linear model, especially when data are scarce, whereas at larger sample sizes the comparison is model- and dimension-dependent. Insurance-specific fine-tuning further improves the embeddings, and a prompt-sensitivity diagnostic shows that the pipeline reacts to any appended out-of-template field, making controlled prompts a governance requirement.

\end{abstract}

\textbf{Keywords}: Insurance pricing, risk classification, underwriting, representation learning, large language models, embeddings

\section{Introduction}

In motor insurance pricing, actuaries use variables such as driver age, vehicle characteristics, geography, and claim history to estimate claim frequency. In non-life insurance, a range of predictive models is used for this purpose, including generalized linear models (GLMs), tree-based and boosting methods, neural networks, and hybrid architectures such as the Combined Actuarial Neural Network of \cite{wuthrich2019cann}; GLMs remain attractive in practice for their transparency and ease of governance. In this paper, we focus on claim frequency. A standard actuarial approach is to fit a Poisson GLM that relates claim frequency to structured covariates such as driver age, vehicle type, region, and bonus-malus level through a log-link function. Because GLMs assume a linear relationship between the covariates and the log-expected claim count, capturing interactions between risk factors requires explicit feature engineering: the actuary must identify which pairs or groups of variables interact and construct the appropriate interaction terms. A GLM without interaction terms may understate risks that appear only in combinations: an 18-year-old driver with a high-powered vehicle in a dense urban area may present a risk profile that is not well represented by separate additive effects for age, vehicle power, and location. Flexible learners such as gradient-boosted trees or neural networks can learn such patterns from data when sufficient observations are available, see \citet{holvoet2025neural} for a comparison study. The question we study is whether an embedding representation can make similar structure usable inside a GLM, especially at small sample sizes.

Recent actuarial work has replaced one-hot or hand-crafted variables with learned representations of categories, locations, text, and images. Embedding-based representations of high-cardinality categorical covariates were introduced to actuaries by \citet{richman2020ai, richman2020aia}. These ideas have been extended in \citep{shi2023nonlife, delong2023use, avanzi2024machine, richman2024highcardinality}. However, these approaches require neural networks to be trained jointly with the pricing model, which may be challenging when the dataset is small or the signal-to-noise ratio is low. The representation learning framework of \citet{blier-wong2021rethinking} decouples the feature construction step from the prediction step: a representation model transforms raw data into dense vectorial embeddings, and a standard GLM then uses these embeddings as features. This framework has been applied to incorporate geographic data \citep{blier-wong2022geographic,holvoet2025multiview}, textual claims data \citep{xu2022bertbased}, and image data \citep{blier-wong2024representationlearning} into insurance pricing models. In these applications, embeddings compress sparse inputs, such as geographic grids, claim notes, or images, into fixed-length vectors that can be used in a GLM.

In parallel, large language models (LLMs) built on the transformer architecture of \cite{vaswani2017attention} produce internal representations that reflect semantic relationships between concepts, including factual knowledge, domain-specific associations, and contextual dependencies learned from massive text corpora \citep{bengio2013representation, reimers2019sentencebert}.
Recent tabular-LLM studies serialize rows as text and often report the largest gains when labeled samples are scarce (see, e.g., \citep{hegselmann2023tabllm, wang2024unipredict}). A related line of work extracts embeddings from LLMs applied to serialized tabular data and uses these as features for a separate supervised model, separating representation from prediction (see, e.g., \citep{villaloboscarballo2022tabtext, koloski2025llmembeddings, kasneci2024enriching}). We review this literature in detail in Section~\ref{ss:related}.

In this paper, we propose using pre-trained LLM embeddings to replace traditional hand-crafted features in the actuarial pricing pipeline. We use the embedding model as a fixed reparameterization of the variables: the resulting vector contains no new fields, but may make some interactions easier for a generalized linear model to use. We do not manually specify any interactions; instead, we test whether embeddings of the serialized policyholder descriptions contain directions that improve a downstream GLM, and we fine-tune the embedding model using low-rank adaptation (LoRA, \cite{hu2022lora}) and a ranking-based contrastive objective that aligns the embedding geometry with claim frequency similarity.

Two properties of the representation motivate this. First, the embedding model maps categorical covariates such as region and vehicle brand to dense vectors that may reflect similarities among category labels rather than treating each level as an isolated one-hot indicator. Second, because self-attention combines information across token positions, the embedding can depend on combinations of fields and not only on each field separately. This motivates testing whether the embeddings carry interaction-relevant variation that a GLM could use; whether either property is useful for claim-frequency pricing is an empirical question.

We evaluate the approach across training set sizes from a few hundred to several hundred thousand policies, using a Poisson GLM with exposure offset as the downstream model, and we fine-tune the embedding model on an insurance-specific task with LoRA; Section~\ref{ss:related} positions these contributions relative to prior work. As a supplementary check (Appendix~\ref{s:fairness}), we also probe whether the embedding pipeline reacts to tokens that should not influence pricing.

The remainder of this paper is structured as follows. Section~\ref{s:background} provides the technical background on transformer architectures, embedding models, fine-tuning, and the connection between embeddings and actuarial pricing. It also reviews related work. Section~\ref{s:methodology} describes the prompt design, embedding extraction, and downstream pricing model. Section~\ref{s:data} presents the data and experimental setup. Sections~\ref{s:results} and~\ref{s:finetuning-experiments} report the off-the-shelf and fine-tuned embedding results, respectively, and Section~\ref{s:discussion} discusses practical considerations and limitations and concludes. We also prepare two appendices to further interpret how the methodology works. Appendix~\ref{s:fairness} a protected-attribute prompt sensitivity diagnostic, and Appendix~\ref{app:clustering} an unsupervised clustering of the embedding into rating bands.

\section{Background}\label{s:background}

\subsection{The transformer architecture}\label{ss:transformers}

Modern large language models are built on the transformer architecture \citep{vaswani2017attention}, which converts an input string into a sequence of tokens and refines a vector representation of each token through a stack of self-attention and feed-forward layers.  The property relevant here is that self-attention lets each token's representation depend on the other tokens in the input, so a field such as a driver's age can in principle be encoded differently depending on the vehicle and location fields that accompany it.  A GLM with additive terms cannot represent such compounded structure unless the relevant interaction terms are supplied explicitly; whether the embedding makes this structure usable for claim-frequency pricing is an empirical question that the experiments below address.  In our pipeline, the contextualized token representations become pricing covariates only after pooling, dimensionality reduction, and GLM fitting.

Modern LLMs are trained on massive text corpora, often comprising trillions of tokens drawn from books, web pages, scientific articles, and code.  They use a next-token prediction objective: at each position, the model learns to predict the next token given all preceding tokens.  This objective encourages representations that encode syntax, semantics, and statistical associations useful for predicting text. Alongside their LLMs for generative modelling, Gemini and Qwen \citep{lee2025gemini, zhang2025qwen3} also provide general-purpose embedding models built from large foundation models and trained across diverse tasks, languages, and domains. We test whether those representations contain structure that is useful for claim-frequency pricing.

\subsection{Text embedding models}\label{ss:embedding-models}

\begin{figure}[ht]
  \centering
  \begin{tikzpicture}[
    node distance=0.45cm,
    stage/.style={draw, rounded corners=2pt, fill=gray!10, align=center, minimum height=0.85cm, inner sep=4pt, font=\small}
  ]
    \node[stage] (cov) {structured\\covariates $\mathbf{x}_i$};
    \node[stage, right=of cov] (prompt) {serialized\\prompt $p_i$};
    \node[stage, right=of prompt] (model) {embedding\\model};
    \node[stage, right=of model] (pool) {pooled embedding\\$\mathbf{e}_i \in \mathbb{R}^d$};
    \node[stage, right=of pool] (pca) {PCA\\$\mathbf{z}_i \in \mathbb{R}^K$};
    \node[stage, right=of pca] (glm) {Poisson\\GLM};
    \draw[arrow] (cov) -- (prompt);
    \draw[arrow] (prompt) -- (model);
    \draw[arrow] (model) -- (pool);
    \draw[arrow] (pool) -- (pca);
    \draw[arrow] (pca) -- (glm);
  \end{tikzpicture}
  \caption{Overview of the embedding-based pricing pipeline: structured covariates are serialized into a prompt, encoded by the embedding model, pooled into a fixed-length vector, reduced by PCA, and used as covariates in a Poisson GLM with exposure offset.}
  \label{fig:pipeline}
\end{figure}
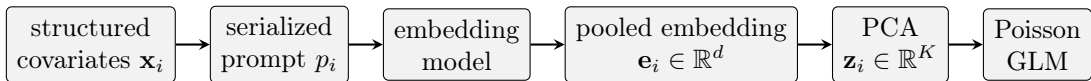

A generative LLM produces text by sampling one token at a time.  An embedding model instead maps an entire input sequence to a single fixed-length vector $\mathbf{e} \in \mathbb{R}^d$, which we use as the feature representation for downstream pricing.  The vector is obtained by pooling the model's final-layer token representations; common choices are mean pooling over token positions, classification-token pooling, and latent attention pooling \citep{lee2025nvembed}.  The choice of pooling affects embedding quality, but mean pooling is the most common default.  Figure~\ref{fig:pipeline} summarizes how the pooled embedding fits into the pricing pipeline studied in this paper.

A raw embedding from a generative LLM that was trained only on next-token prediction may not produce vectors that are well-suited for measuring semantic similarity, because the training objective does not explicitly encourage similar inputs to have similar representations.  Dedicated embedding models address this through additional training stages.  The typical pipeline, as described by \citet{wang2022e5, zhang2025qwen3}, proceeds in two or three stages.  In the first stage, the model is pre-trained on a large corpus of text pairs using a contrastive objective that encourages matched pairs, such as a query and its relevant passage, to have high cosine similarity while pushing unrelated pairs apart.  In the second stage, the model is fine-tuned on a curated collection of supervised tasks, including semantic similarity, retrieval, classification, and clustering, to improve generalization across diverse embedding applications.  Some models add a third stage that merges weights from multiple fine-tuning runs to balance performance across tasks.

The result is a model that maps any text input to a dense vector in $\mathbb{R}^d$, where cosine similarity between vectors reflects semantic relatedness.  The dimension $d$ varies across models: Qwen3-Embedding-0.6B produces vectors in $\mathbb{R}^{1024}$, the Qwen3-Embedding-8B variant in $\mathbb{R}^{4096}$, Gemini Embedding in $\mathbb{R}^{3072}$, and Llama-Embed-Nemotron-8B in $\mathbb{R}^{4096}$.  Recent models support Matryoshka representations \citep{kusupati2022matryoshka}, in which the first $d'$ coordinates of the full $d$-dimensional embedding remain useful at any truncation level $1 \leq d' < d$, enabling flexible dimensionality reduction without retraining.  Some embedding models are instruction-aware \citep{su2023instructor}: they accept a task description alongside the input text through a dedicated interface, enabling the same model to produce embeddings tailored to different downstream tasks; for instance, one may supply an instruction such as ``Represent this insurance policyholder description for risk classification'' to guide the model toward risk-relevant features of the input.  For models without such an interface, task text can instead be included directly in the input prompt, where it is processed as ordinary input; our prompt templates in Section~\ref{s:methodology} and Appendix~\ref{app:prompts} use this latter mechanism for all models, including the instruction-aware Qwen3-Embedding family. This choice keeps the serialization identical across embedding models, but it means the Qwen3-Embedding results are not directly comparable with that model's benchmark configuration. 

On the Massive Multilingual Text Embedding Benchmark (MMTEB; \citep{enevoldsen2025mmteb}), Qwen3-Embedding \citep{zhang2025qwen3} and Gemini Embedding \citep{lee2025gemini} rank competitively.  Our experiments use Qwen3-Embedding and Gemini Embedding, together with Llama-Embed-Nemotron-8B \citep{babakhin2025llamaembednemotron}, chosen to span proprietary and open-source deployment, a range of embedding dimensions, and local fine-tuning feasibility.  The Qwen3-Embedding family is available in several variants, including 0.6B- and 8B-parameter models, with the smaller model suitable for local deployment and fine-tuning on a single graphics processing unit (GPU). We refer to \citet{neelakantan2022text} for an early demonstration that contrastive pretraining at scale and larger model backbones can improve text-embedding quality, and to \citet{lee2024gecko} for an approach that distills knowledge from a large LLM into a compact embedding model.

\subsection{Fine-tuning embedding models}\label{ss:finetuning}

Off-the-shelf embedding models are trained on general-purpose text corpora and optimized for tasks such as web search and document retrieval.  For example, a general embedding model may represent ``diesel'' and ``gasoline'' as related fuel types, but may not encode the actuarial meaning of a Bonus-Malus score of 150 versus 50.  Fine-tuning adapts the embedding geometry so that the distance between two policyholder embeddings reflects their similarity in terms of insurance risk rather than general linguistic meaning.

Fine-tuning a model with hundreds of millions or billions of parameters on a moderately-sized insurance dataset poses a computational challenge because updating all parameters is computationally expensive and prone to overfit.  Low-Rank Adaptation (LoRA; \citep{hu2022lora}) addresses the computational and memory cost of full fine-tuning by freezing all pre-trained weights and injecting small trainable matrices into each transformer layer.  Specifically, for a pre-trained weight matrix $\mathbf{W}_0 \in \mathbb{R}^{d_1 \times d_2}$, LoRA parameterizes the update as
\begin{equation}\label{eq:lora}
  \mathbf{W} = \mathbf{W}_0 + \frac{\alpha}{r} \, \mathbf{B} \mathbf{A},
\end{equation}
where $\mathbf{A} \in \mathbb{R}^{r \times d_2}$ and $\mathbf{B} \in \mathbb{R}^{d_1 \times r}$ are trainable low-rank matrices with $r \ll \min(d_1, d_2)$, and $\alpha > 0$ is the LoRA scaling parameter.  Only $\mathbf{A}$ and $\mathbf{B}$ are updated during training, reducing the number of trainable parameters by a factor of $d_1 d_2 / (r(d_1 + d_2))$; the reference rank and the resulting fraction of trainable parameters are given in Section~\ref{ss:finetuning-procedure}.  Because $\mathbf{W}_0$ remains frozen, adaptation is confined to the low-rank update $\mathbf{B}\mathbf{A}$, which adjusts the embedding space to reflect insurance-specific risk relationships. The LoRA procedure has become the standard method to fine-tune LLMs to more narrow tasks when the downstream task does not have much training data. 

The fine-tuning objective determines what structure the adapted embedding space will have.  Since our goal is to produce embeddings in which policyholders with similar claim frequencies lie close together, we use the CoSENT loss \citep{huang2024cosent}, which operates on pairs of inputs.  Given a batch of policyholder prompts $\{p_i\}_{i=1}^n$ with observed claim frequencies $f_i=N_i/E_i$ and $E_i>0$, we first define a pairwise similarity score
\begin{equation}\label{eq:similarity}
  s_{ij} = \exp\!\left\{-\frac{|\log(1+f_i)-\log(1+f_j)|}{\tau}\right\} \in (0, 1],
\end{equation}
where $\tau > 0$ controls sensitivity to log-frequency differences: pairs with identical frequencies receive a similarity of 1, while pairs with large log-frequency differences receive a similarity near 0.  The CoSENT loss then encourages the cosine similarity between embeddings $\mathbf{e}_i$ and $\mathbf{e}_j$ to respect the ranking induced by $s_{ij}$.  Formally, for pairs $(i,j)$ and $(k,l)$ with $s_{ij} > s_{kl}$, the loss penalizes configurations where the embedding cosine similarities violate this ordering:
\begin{equation}\label{eq:cosent}
  \mathcal{L}_{\mathrm{CoSENT}}\!\left(\{\mathbf{e}_i\}_{i=1}^n,\{s_{ij}\};\lambda\right)
  = \log\left(1 + \sum_{\substack{(i,j),(k,l) \\ s_{ij} > s_{kl}}}
  \exp\!\left\{\lambda\bigl(\cos(\mathbf{e}_k, \mathbf{e}_l) - \cos(\mathbf{e}_i, \mathbf{e}_j)\bigr)\right\}\right).
\end{equation}
This ranking-based formulation is more robust than losses that target exact cosine similarity values, because it only requires the model to preserve the relative ordering of risk similarity among pairs.  The scale parameter $\lambda > 0$ controls the sharpness of the ranking penalty; following the CoSENT experiments in \citet{huang2024cosent}, we use $\lambda = 20$.  The scale $\lambda$ and the similarity temperature $\tau$ of \eqref{eq:similarity} are fixed across all fine-tuning runs to the values given in Section~\ref{ss:finetuning-procedure}.

The combination of LoRA and CoSENT enables us to fine-tune the embedding model efficiently.  Training updates fewer than 1\% of the parameters, and the adapted embedding geometry aligns with the actuarial objective of grouping similar risks together and separating dissimilar ones.  After fine-tuning, the model produces an embedding function $f_{\text{embed}}: \mathcal{T} \to \mathbb{R}^d$, where $\mathcal{T}$ denotes the set of text prompts, that maps any policyholder prompt $p_i$ to a $d$-dimensional vector whose geometry reflects claim frequency similarity.

\subsection{From embeddings to insurance pricing}\label{ss:embeddings-to-pricing}

In \citet{blier-wong2021rethinking}, the insurance pricing process is decomposed into two steps: a representation model that transforms raw data into a feature vector, and a predictive model that maps this feature vector to an estimate of the expected loss.

Consider a portfolio of $n$ insurance contracts, where each contract $i \in \{1, \dots, b\}$ is described by a vector of covariates $\mathbf{x}_i$ (driver age, vehicle type, region, Bonus-Malus score, etc.), an exposure $E_i$, and a claim count $N_i$.  We assume $E_i>0$ and model $N_i$ conditional on covariates and exposure as a Poisson distributed random variable with mean $\mu_i=E_i\lambda_i$; the offset formulation estimates the conditional claim frequency $\lambda_i$.  The standard actuarial approach is to fit a Poisson GLM of the form
\begin{equation}\label{eq:glm-baseline}
  \ln(\mu_i) = \ln(E_i) + \beta_0 + \boldsymbol{\beta}^\top \mathbf{x}_i^*,
\end{equation}
where $\mu_i = \mathbb{E}[N_i \mid \mathbf{x}_i,E_i]$, $\mathbf{x}_i^*$ is a vector of hand-engineered features derived from $\mathbf{x}_i$, including one-hot encodings of categorical variables, polynomial terms, and selected interactions, and $\boldsymbol{\beta}$ is the vector of regression coefficients estimated by maximum likelihood.

Our approach replaces $\mathbf{x}_i^*$ with features derived from LLM embeddings.  We first serialize the covariate vector $\mathbf{x}_i$ into a natural language prompt $p_i = T(\mathbf{x}_i)$ using a deterministic template function $T$ (described in Section~\ref{s:methodology}).  We then compute the embedding $\mathbf{e}_i = f_{\text{embed}}(p_i) \in \mathbb{R}^d$.  Because the embedding dimension $d$ is large relative to the desired number of GLM coefficients, and because unregularized high-dimensional GLMs can be unstable at small sample sizes, we apply principal component analysis (PCA) to reduce the dimensionality.  The reduced features are
\begin{equation}\label{eq:pca}
  \mathbf{z}_i = \text{PCA}_K(\mathbf{e}_i) \in \mathbb{R}^K,
\end{equation}
where $K \geq 1$ is a tuning parameter.  The pricing model then becomes
\begin{equation}\label{eq:glm-embedding}
  \ln(\mu_i) = \ln(E_i) + \gamma_0 + \boldsymbol{\gamma}^\top \mathbf{z}_i,
\end{equation}
where $\mu_i = \mathbb{E}[N_i \mid \mathbf{z}_i,E_i]$ and $\boldsymbol{\gamma} \in \mathbb{R}^K$ is estimated by maximum likelihood as before.  The downstream model remains a Poisson GLM fitted by maximum likelihood.

The LLM serves only as a feature extractor: it outputs embeddings and never a premium, and the fitted claim frequency is produced by the GLM.  Because embedding extraction is a single deterministic forward pass with no token sampling, a given description always yields the same embedding and the same prediction, which rules out the hallucination and run-to-run variability that arise when a generative model is asked to produce a number directly.  This does not remove the representation or bias risks inherited from the embedding model, as the diagnostic in Appendix~\ref{s:fairness} shows, but removes the hallucination issue of LLMs. 

The representation and prediction steps are modular: the actuary may swap the embedding model, adjust the PCA dimension, or replace the GLM with a gradient boosting machine or neural network without modifying the other components.  This is consistent with the representation learning framework of \citet{blier-wong2021rethinking}, where the same embeddings may be reused across different pricing models and lines of business.

We test whether LLM embeddings expose relationships between covariates, that is, interactions and non-linear effects, that an additive GLM cannot exploit from the raw covariates unless suitable transformations or interaction terms are supplied.  The raw covariates contain this information; the question is whether the embedding makes it accessible to the downstream model.  If this hypothesis is correct, then even a simple linear model operating on the embedding features should outperform the same model operating on raw or hand-crafted features, particularly in settings where the training data are insufficient for the model to learn these relationships from the response variable alone.

\subsection{Related work}\label{ss:related}

We organize the related work into three streams.  The first concerns LLMs applied to tabular prediction through text serialization.  The second, which is most closely related to our contribution, concerns extracting LLM embeddings as features for downstream supervised models.  The third covers embeddings and natural language processing in actuarial science.

\subsubsection{LLMs for tabular prediction via serialization}

Early examples of converting a tabular data row into natural language and processing it with a language model include LIFT \citep{dinh2022lift} and TabLLM \citep{hegselmann2023tabllm}.  The LIFT approach serializes each sample into a sentence of the form ``When we have $x_1 = r_1, x_2 = r_2, \ldots$, what should be $y$?'' and fine-tunes pretrained language models with cross-entropy loss; the approach requires no architectural modifications and produces well-calibrated predictions.  The TabLLM study evaluated nine serialization formats, ranging from simple text templates to LLM-generated prose, for few-shot classification, finding that simple text templates of the form ``The [feature name] is [value]'' consistently outperform more elaborate serializations, including those generated by GPT-3, which suffer from hallucination artifacts.  These papers place language-model-based tabular prediction in low-data regimes: TabLLM is strongest in zero- and very-few-shot settings and is often competitive with XGBoost and TabPFN as the number of shots increases.

Subsequent work has examined the serialization step more carefully.  In \citet{jaitly2023text}, the authors study how text serialization interacts with conventional tabular-learning practices and find that language-model performance is sensitive to serialization and data-curation choices, with feature selection appearing more consistently useful than feature scaling or missing-value imputation.  A survey by \citet{fang2024llmtabular} documents the effect of serialization format, including text, JavaScript Object Notation, HyperText Markup Language, and comma-separated values, on prediction, generation, and table understanding tasks across a wide range of models and datasets.  These studies suggest that simple text templates are often competitive, especially when labels are scarce. The evidence is less clear about which parts of the prompt the model uses.

Other papers train or adapt LLMs directly for end-to-end tabular prediction.  The UniPredict model \citep{wang2024unipredict} trains a single GPT-2 on 169 tabular datasets via instruction tuning, serializing feature names and values into natural language with metadata; the resulting model outperforms the best per-dataset baselines, with the largest advantage in low-resource settings.  The Tabula approach \citep{gardner2024tabula} demonstrates large-scale transfer learning for tabular data via language modeling, fine-tuning LLaMA on a web-scale corpus of millions of tables and evaluating transfer on 329 tabular datasets.  In \citet{wen2024supervised}, the authors propose a generative paradigm for tabular deep learning, using LLMs to model the joint distribution of features and labels.  An evaluation by \citet{yang2025unleashing} identifies conditions under which LLM-based approaches outperform conventional methods.  In all of these works, the low-data regime is where LLMs provide the largest gains, consistent with the view that the semantic knowledge embedded in pre-trained models acts as an effective prior when labeled observations are scarce.

Another approach uses LLMs for feature engineering.  The CAAFE method \citep{hollmann2023caafe} iteratively queries an LLM with dataset descriptions to generate Python code for new features, improving mean classification performance across 14 datasets.  The FeatLLM approach \citep{han2024featllm} uses LLMs to generate rules rather than code, producing interpretable binary features.  In \citet{abhyankar2025llmfe}, the authors formulate feature engineering as an evolutionary program search, using LLMs as knowledge-guided optimizers.  These approaches produce interpretable features but are limited to the transformations that the LLM can explicitly articulate; as we discuss next, extracting dense latent representations from LLMs may capture richer information than explicit rules.

A broader line of work builds purpose-built tabular foundation models, which are a useful boundary case for our contribution because they intervene at a different level of the modeling pipeline.  The TabPFN model \citep{hollmann2025tabpfn} is a transformer pre-trained across millions of synthetic tabular datasets; at inference it consumes the labeled training rows and the test covariates together as context and infers labels in a single forward pass, outperforming previous methods on datasets with up to 10,000~samples.  TabPFN therefore learns a tabular prediction algorithm rather than a representation: it approximates the posterior predictive distribution $p(y \mid \mathbf{x}, D)$ for a test point $\mathbf{x}$ given a labeled dataset $D$, amortizing Bayesian-style prediction over both the dataset and the test point, and so replaces the fitted GLM, tree, or boosting model with an in-context predictor.  The TabDPT approach \citep{ma2025tabdpt} scales this paradigm to real data.  Related tabular architectures include TabTransformer \citep{huang2020tabtransformer} and FT-Transformer \citep{gorishniy2021revisiting}, which use self-attention over feature embeddings within a table; TransTab \citep{wang2022transtab}, which learns transferable tabular transformers across tables; and CARTE \citep{kim2024carte}, which supports pretraining and transfer across tables using graph-attentional contextualization of table entries and column names.

\subsubsection{Extracting LLM embeddings for downstream tabular models}

The approach most closely related to our contribution extracts the internal vector representations, that is, embeddings, produced by LLMs when processing serialized tabular data.  These embeddings then serve as input features for a separate supervised model such as a gradient-boosted tree or a GLM.  This two-step pipeline separates the representation model from the prediction model, a decomposition that has both practical and theoretical advantages. The LLM serves only as a representation model, avoiding generative prediction and the associated hallucination concerns; the downstream model can be analyzed using the tools appropriate to its model class, conditional on the constructed features; and the embeddings may be precomputed and reused across different prediction tasks.

The TabText method \citep{villaloboscarballo2022tabtext} is an early example of this idea.  They serialize tabular medical records into contextual language using an ``attribute: value'' format, concatenating fields into patient descriptions, and apply pre-trained clinical language models (BioGPT, Clinical-Longformer) to generate task-independent embeddings.  These fixed embeddings serve as inputs for gradient-boosted trees, linear models, and support vector machines.  On nine inpatient flow prediction tasks, augmenting preprocessed tabular features with TabText embeddings improved predictive performance and demonstrated cross-institution generalization, suggesting that LLM-derived representations transfer across institutional contexts.

In \citet{koloski2025llmembeddings}, the authors evaluate the approach across seven classification benchmarks.  They serialize features individually (``This [feature name] is [value]''), extract embeddings from frozen LLMs (BGE, LLaMA-3, MiniLM), project through a shallow multilayer perceptron, and feed the result into trainable architectures including a multilayer perceptron, ResNet, and FT-Transformer.  The FT-Transformer gained accuracy from LLM embeddings, and the BGE model produced the most consistent improvements.  Datasets with prominent categorical features benefited most, while highly domain-specific features poorly covered by the LLM's pre-training corpus showed degradation, a finding relevant to our insurance application, where domain-specific fine-tuning may be necessary.

In \citet{kasneci2024enriching}, the authors conduct an ablation study in which tabular rows are serialized to text, embedded with RoBERTa and GPT-2, reduced to 50 principal components via PCA, and combined with the original features for use in Random Forest, XGBoost, and CatBoost.  The LLM-derived features frequently ranked among the most important in feature importance analyses, and XGBoost and CatBoost benefited most from the augmented feature set.  Similar findings appear in \citep{haque2025leveraging} in a graduate employment classification task, where LLM-based feature selection and embedding improved downstream model performance by providing richer representations of structured covariates.

Two recent papers address the geometry of LLM embeddings for tabular tasks.  The LATTE method \citep{shi2025latte} extracts task-relevant knowledge from LLaMA-2 hidden states via task metadata prompts and distills this knowledge into a tabular encoder using Kullback-Leibler divergence.  They find that latent-level knowledge transfer, that is, transferring the embedding representation itself, outperforms text-level feature engineering, suggesting that LLM hidden states encode predictive information that explicit rules cannot fully articulate.  This observation motivates our approach of fine-tuning the embedding geometry rather than generating explicit features.  In \citet{kang2026tarl}, the authors find that raw LLM embeddings of serialized table rows contain useful predictive signal but that naive application underperforms due to embedding anisotropy, the tendency of LLM embeddings to cluster in a narrow cone of the vector space.
Their lightweight fix, Common Component Removal with per-task temperature calibration, achieves competitive performance with specialized tabular foundation models on semantically rich data while providing a substantial speedup over the standard approaches. 

Prior work shows that serialized-table embeddings can help downstream learners, but most of this evidence concerns classification tasks, generic tabular benchmarks, or flexible learners.  The open question for insurance pricing is narrower: whether an embedding representation improves a constrained Poisson GLM used for claim-frequency pricing.  Our approach differs from prior work in three respects.  First, the downstream model is a Poisson GLM with exposure offset, a constrained model class that cannot learn interactions on its own and so may benefit from pre-computed interaction-relevant representations.  Second, we fine-tune the embedding model on an insurance-specific task with a CoSENT loss that aligns embedding geometry with claim-frequency similarity, rather than using frozen general-purpose embeddings.  Third, we evaluate across training set sizes from a few hundred to several hundred thousand policies.

\subsubsection{Embeddings and natural language processing in actuarial science}

The use of embeddings in actuarial science predates the current wave of LLM research. \citet{richman2020ai, richman2020aia} introduced actuarial readers to embedding-based supervised representations for high-cardinality categorical variables and illustrated their use in insurance modelling. \citet{shi2023nonlife} then provided an in-depth non-life case study, showing that categorical embeddings may be used for risk classification, transfer learning across products, and improved treatment of rare categories. These ideas have been extended in subsequent work on high-cardinality covariates \citep{richman2024highcardinality, avanzi2024machine} and neural-network training with mixed categorical and numerical features \citep{delong2023use, holvoet2025neural}. The common thread is that learned embeddings capture relationships between categories, such as the similarity between occupations or vehicle types, that one-hot encoding discards.

The representation learning framework for insurance pricing was formalized by \citet{blier-wong2021rethinking}, who decompose the ratemaking process into two steps: a representation model that transforms raw data into dense vectorial embeddings, and a regression model (typically a GLM) that uses these embeddings as features.  This framework was applied to geographic data using convolutional autoencoders and spatial embeddings \citep{blier-wong2022geographic, holvoet2025multiview}, to textual claims data using BERT-based encoders \citep{xu2022bertbased}, and to images of insured dwellings using transfer learning from pre-trained convolutional neural networks \citep{blier-wong2024representationlearning}.  In each case, the unstructured data source is high-dimensional and sparse, the representation model compresses it into a fixed-length vector, and the downstream GLM operates on these compressed features.  The present paper applies this same two-step methodology to structured covariates themselves.  We transform them into natural language and embed them using a pre-trained LLM, producing representations that may make contextual and interaction-relevant structure, already implicit in the raw covariates, accessible to a downstream GLM.

Natural language processing (NLP) has been applied directly to insurance text data in several settings.  In \citet{lee2020actuarial}, the authors extract word embeddings from actuarial text corpora and use cosine similarities between claim descriptions for classification and loss prediction.  \citet{baillargeon2020mining} use recurrent neural networks to extract textual predictors from accident descriptions and estimate the number of cars involved in an accident within a Poisson regression model.  The work of \citet{xu2022bertbased} extracts BERT embeddings from claim text and shows that the NLP component, not just the neural network architecture, drives improvements in classification and severity modeling in warranty data.  In \citet{campo2024clustering}, the authors use word embeddings to cluster industry codes and worker's compensation classes into homogeneous risk groups, showing that semantic similarity between textual descriptions of categories serves as a useful proxy for actuarial similarity.  See also \citep{zappa2021text, dong2025insurtecha} for other applications of text mining in insurance, and \citep{troxler2022actuarial, balona2024actuarygpt, hatzesberger2025advanced, balona2025operationalizing} for overviews of generative artificial intelligence applications in actuarial practice.

\section{Methodology}\label{s:methodology}

\subsection{Input serialization}

The first step of our pipeline converts each policyholder's structured covariate vector into a natural-language prompt that an embedding model can process.  The prompt must include all variables and express them with labels and scales that the embedding model can parse.  For this reason, we adopt an instruction-prefixed underwriting list as the default template.  The template provides contextual anchors for numeric fields, such as the entrance-level Bonus-Malus score and the average population density in the insurance portfolio.  The qualitative binning thresholds of Appendix~\ref{app:prompt-qualitative} are likewise fixed domain constants chosen from actuarial knowledge, and the standardization, PCA loadings, and downstream GLM are fitted on the training subset only.  The prompt also labels each covariate with a human-readable description rather than an abbreviated variable name.  Figure~\ref{fig:prompt-baseline} displays the template used throughout the main experiments; we compare this default prompt against five alternative serialization formats in Section~\ref{sec:prompt-format}.

\begin{figure}[ht]
  \centering
  \begin{prompt}
You are an auto insurance underwriter. Evaluate the risk level of a policyholder based strictly on the following insurance-related information:
- Policyholder Age: {DrivAge} years old (in France, people can drive starting at age 18)
- Land Type: {Area}
- Region: {Region}, France
- Population density: {Density} people/km2 (average density in this insurance dataset is 1792 people/km2)
- Vehicle: {VehBrand}
- Vehicle Age: {VehAge} years old
- Fuel type: {VehGas} (either Diesel or Regular Gasoline)
- Power class: {VehPower} (min = 4, max = 15)
- Bonus-Malus score: {BonusMalus} (scored between 50 and 230 with entrance level 100, <100 means bonus, >100 means malus)
\end{prompt}
\caption{Default instruction-prefixed underwriting-list prompt for generating LLM embeddings from structured features. Each placeholder (e.g., \texttt{\{DrivAge\}}) is replaced with the corresponding feature value for each policyholder.}
\label{fig:prompt-baseline}
\end{figure}

The prompt template is fixed before model fitting and is treated as part of the preprocessing pipeline: the prompt exposes all variables to the embedding model in a labelled, insurance-specific form, and the downstream GLM remains responsible for estimating claim frequency.  The instruction and contextual anchors shape the representation geometry; they do not ask the LLM to produce a premium or a claims prediction.

\subsection{Embedding extraction}

We compare four embedding models: Google Gemini Embedding \citep{lee2025gemini}, Qwen3-Embedding-0.6B \citep{zhang2025qwen3}, Qwen3-Embedding-8B \citep{zhang2025qwen3}, and Llama-Embed-Nemotron-8B \citep{babakhin2025llamaembednemotron}; in tables and figures these are abbreviated Gemini, Qwen3-0.6B (or Qwen), Qwen3-8B, and Nemotron-8B, respectively.  These models span proprietary and open-source deployment, a range of embedding dimensions, and differing local fine-tuning feasibility.  The Qwen3-Embedding-0.6B model is open source and small enough for local deployment and fine-tuning on a single GPU.  The Qwen3-Embedding-8B and Llama-Embed-Nemotron models require more substantial compute; they serve as benchmarks for whether larger open-source models yield better embeddings for our task.  The Gemini model is accessed through an application programming interface, which may limit customization but provides access to a large proprietary model.

Formally, let $f_{\text{embed}} : \mathcal{T} \to \mathbb{R}^d$ denote the embedding model, which maps a text prompt to a $d$-dimensional vector.  For each policyholder $i$, we compute
\begin{equation}\label{eq:embedding}
  \mathbf{e}_i = f_{\text{embed}}(p_i) \in \mathbb{R}^d.
\end{equation}
The embedding dimension $d$ depends on the model: Gemini Embedding produces vectors in $\mathbb{R}^{3072}$, Qwen3-Embedding-0.6B in $\mathbb{R}^{1024}$, Qwen3-Embedding-8B in $\mathbb{R}^{4096}$, and Llama-Embed-Nemotron-8B in $\mathbb{R}^{4096}$.  Because $d$ is large relative to the number of raw features from the GLM baseline, and because unregularized high-dimensional GLMs can be unstable at small training sizes, we reduce the dimensionality before fitting the downstream GLM.  One candidate is Matryoshka truncation \citep{kusupati2022matryoshka}, which trains the model so that the first $d'$ coordinates of the full $d$-dimensional embedding remain useful at any truncation level $d' < d$.  However, Matryoshka representations are optimized for general-purpose retrieval and similarity tasks across diverse text inputs; the ordering of dimensions reflects what is informative for distinguishing arbitrary text passages, not what is informative for distinguishing insurance policyholders.

Because the prompts share the same template and domain, much of the embedding variation may reflect common prompt text rather than policyholder differences.  The raw embeddings occupy a narrow region of $\mathbb{R}^d$: a large portion of each embedding encodes information that is constant across our prompts, such as the fact that the text describes a person, a vehicle, and a geographic location in a motor insurance context.  This shared component is uninformative for pricing because it does not vary across policyholders.  What matters for the downstream GLM are the directions along which policyholder embeddings differ from one another, and these directions need not align with the leading Matryoshka coordinates.

We therefore apply principal component analysis (PCA) to the $n \times d$ embedding matrix $\mathbf{E} = [\mathbf{e}_1, \ldots, \mathbf{e}_n]^\top$, retaining the first $K$ components.  For each downstream training size, embeddings are standardized and PCA is fitted only on the embeddings of that downstream training subset.  The fitted centering/scaling vector and loading matrix are then applied unchanged to validation or test embeddings.  Thus $\text{PCA}_K$ below denotes the transformation fitted on the selected downstream training subset:
\begin{equation}\label{eq:pca-method}
  \mathbf{z}_i = \text{PCA}_K(\mathbf{e}_i) \in \mathbb{R}^K.
\end{equation}
PCA then retains the directions of largest variation among policyholder embeddings.  These directions need not be the most predictive directions, so the usefulness of the retained PCs is assessed empirically.  This data-driven dimensionality reduction is more appropriate for our application than Matryoshka truncation when the informative directions are not aligned with the native coordinate order.  It is also related to the embedding anisotropy problem identified by \citet{kang2026tarl}, in which LLM embeddings of similar inputs cluster in a narrow cone of the vector space, although PCA should not be interpreted as guaranteeing removal of all anisotropic structure.

In our main experiments, we set $K = 48$ to match the number of downstream GLM coefficients in the baseline model.  This choice does not equalize the complexity of the upstream embedding model or PCA preprocessing.  We study the sensitivity of the results to the choice of $K$ in Section~\ref{s:results}.

\subsection{Downstream pricing model}\label{ss:downstream-model}

The PCA-reduced embedding vectors $\mathbf{z}_i$ serve as covariates in a Poisson generalized linear model with a log-link function and exposure offset.  As in the baseline model in \eqref{eq:glm-baseline}, we assume $E_i>0$ and model $N_i \mid \mathbf{z}_i,E_i$ as Poisson with mean $\mu_i=E_i\lambda_i$:
\begin{equation}\label{eq:glm-embed}
  \ln(\mu_i) = \ln(E_i) + \gamma_0 + \boldsymbol{\gamma}^\top \mathbf{z}_i,
\end{equation}
where $\mu_i = \mathbb{E}[N_i \mid \mathbf{z}_i,E_i]$ is the expected claim count, $E_i$ is the exposure (treated as an offset), $\gamma_0$ is the intercept, and $\boldsymbol{\gamma} \in \mathbb{R}^K$ is the vector of regression coefficients estimated by maximum likelihood.  The downstream model remains a Poisson GLM fitted by maximum likelihood. The only difference from the baseline GLM in~\eqref{eq:glm-baseline} is that the hand-engineered features $\mathbf{x}_i^*$ are replaced by the data-driven embedding features $\mathbf{z}_i$.

\section{Data and experimental setup}\label{s:data}

We use the French motor third-party liability (MTPL) claims dataset, available through the R package \texttt{CASdatasets} \citep{dutang2020fremtpl} and described in detail in Section 13.1 of \citet{wuthrich2023statistical}.  The dataset contains policy records with their claim counts.  After cleaning, which caps claim counts at five per policy and censors exposures at one calendar year, the cleaned dataset contains 676,782 insurance policies with a total of 35,976 claims.  The overall empirical claim frequency
\[
  \bar{\lambda}=\frac{\sum_{i = 1}^n N_i}{\sum_{i = 1}^n E_i}=\frac{35{,}976}{357{,}134}=10.07\%.
\]

Each policy record contains nine feature components: Area (categorical, ordinal with 6 levels), VehPower (vehicle power class, ordinal/integer with values 4 to 15), VehAge (vehicle age in years), DrivAge (driver age in years), BonusMalus (bonus-malus level between 50 and 230, with entrance level 100), VehBrand (car brand, categorical with 11 levels), VehGas (fuel type, binary: diesel or regular), Density (population density per km$^2$ at the driver's residence), and Region (22 French administrative regions prior to 2016).  The dataset exhibits several characteristics that are common in insurance portfolios.  Claims are uncommon events with a heavily imbalanced response, the BonusMalus score and DrivAge are known to be important predictors of claim frequency in this dataset \citep{wuthrich2023statistical}.

After cleaning, we split the cleaned data at random into three disjoint partitions: a 500,000-policy downstream training pool, a 100,000-policy test set, and a fine-tuning-only pool of the remaining 76,782 cleaned policies, used exclusively for the LoRA experiments of Section \ref{s:finetuning-experiments}.  The fine-tuning-only policies are excluded from both the downstream training pool and the test set before any downstream GLM or XGBoost model is fitted.  To evaluate learning curves, we train models on deterministic subsets of the downstream training pool, ranging from 500 to 500,000 rows, while always evaluating on the same held-out test set.

Our baseline is the Poisson GLM developed in Section 5.2.4 of \citet{wuthrich2023statistical}, which represents a standard actuarial pricing model for this dataset.  The model uses a log-link function with exposure as an offset and includes hand-engineered features: one-hot encodings of categorical variables, polynomial and spline transformations of continuous variables, and selected interactions.  After feature engineering, the baseline GLM uses 48 covariates, which serves as the reference downstream coefficient count when choosing the number of PCA components for the embedding-based models.

We evaluate all models using the mean Poisson deviance on the test set, defined as
\begin{equation}\label{eq:deviance}
  D = \frac{1}{n_{\text{test}}} \sum_{i \in \text{test}} 2 \left( N_i \ln \frac{N_i}{\hat{\mu}_i} - (N_i - \hat{\mu}_i) \right),
\end{equation}
where $N_i$ is the observed claim count, $\hat{\mu}_i = E_i \exp(\hat{\eta}_i)$ is the fitted expected claim count, and the convention $0 \log(0/\hat{\mu}_i)=0$ is used.  Under the log-link model and positive exposure, the fitted means satisfy $\hat{\mu}_i>0$.  In actuarial contexts, even modest reductions in deviance, on the order of 1\%, may be financially substantial when applied across a large portfolio.

\section{Results}\label{s:results}

\subsection{Baseline comparison: raw features vs.\ off-the-shelf embeddings}

We begin by comparing the baseline Poisson GLM fitted on hand-engineered features against Poisson GLMs fitted on PCA-reduced LLM embeddings.  For Gemini, Qwen3-Embedding-0.6B, Qwen3-Embedding-8B, and Llama-Embed-Nemotron-8B, we generate embeddings for all policyholder prompts in the processed train-test split and reduce them to three dimensionalities: $K \in \{12, 48, 192\}$ principal components.  For each training size, PCA is fitted only on the selected downstream training subset and then applied to the test embeddings.  The choice of $K = 48$ matches the baseline GLM's downstream covariate count.  The larger value of $K$ tests whether additional embedding dimensions carry useful signal beyond what 48 components capture.  We train all models on progressively larger subsets of the training data, from 500 to 500,000 rows, evaluating test mean Poisson deviance at each size.

\begin{figure}[ht]
  \centering
  \includegraphics[width=\linewidth]{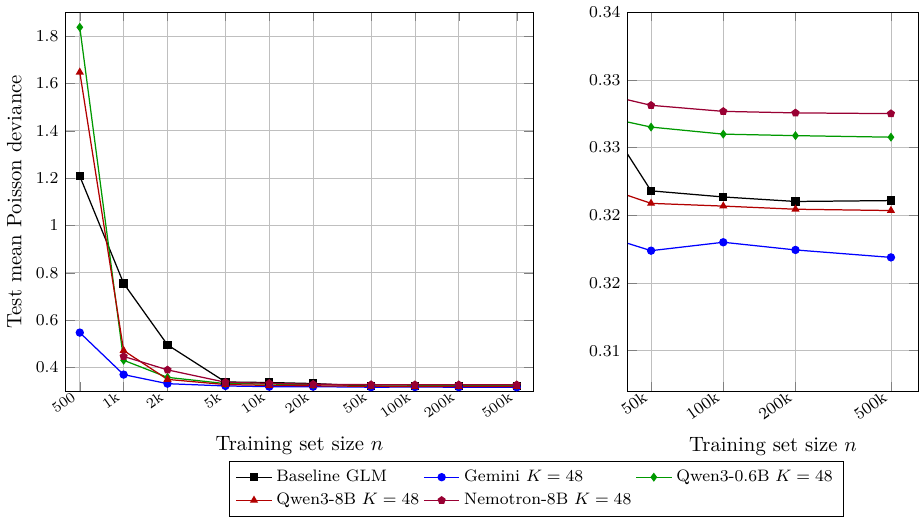}
  \caption{Learning curves comparing the baseline GLM (48 hand-engineered features) against off-the-shelf embedding GLMs with $K = 48$ PCA components.  The right panel zooms in on the large-sample regime.}
  \label{fig:learning-curve-base}
\end{figure}

Table~\ref{tab:baseline-comparison} reports the test deviance at several training set sizes for selected values of $K$.  Figure~\ref{fig:learning-curve-base} displays the corresponding learning curves at $K = 48$, matching the baseline GLM's covariate count.  When only a few hundred observations are available, high-dimensional PCA features are difficult to estimate and the best embeddings are the most compressed ones.  Once the training set reaches a few thousand observations, Gemini provides the strongest low- and medium-sample performance, especially at modest dimension.  With much larger training sets, the higher-dimensional Qwen and Nemotron representations become competitive because the downstream GLM has enough observations to estimate the additional components.  Thus the value of the embedding representation depends on both model and sample size: small samples favour the most compressed representations, whereas large samples may exploit richer embedding coordinates.

\begin{table}[ht]
  \centering
  \caption{Test mean Poisson deviance for the baseline GLM and off-the-shelf embedding GLMs. A dash indicates a non-converged fit or a test deviance greater than 100.}
  \label{tab:baseline-comparison}
  \begin{tabular}{lcccccc}
    \toprule
    Model & $K$ & $n = 500$ & $n = 2{,}000$ & $n = 10{,}000$ & $n = 50{,}000$ & $n = 500{,}000$ \\
    \midrule
    Baseline GLM  & n/a & 1.208 & 0.4961 & 0.3371 & 0.3218 & 0.3211 \\
    \midrule
    \rowcolor{modelgray} Gemini        & 12 & 0.3659 & \textbf{0.3235} & \textbf{0.3180} & 0.3195 & 0.3194 \\
    \rowcolor{modelgray} Gemini        & 48 & 0.5478 & 0.3322 & 0.3197 & 0.3174 & 0.3169 \\
    \rowcolor{modelgray} Gemini        & 192 & -- & 0.6515 & 0.3361 & 0.3192 & 0.3171 \\
    Qwen3-0.6B    & 12 & 0.3944 & 0.3394 & 0.3306 & 0.3300 & 0.3298 \\
    Qwen3-0.6B    & 48 & 1.837 & 0.3589 & 0.3307 & 0.3265 & 0.3258 \\
    Qwen3-0.6B    & 192 & -- & 0.8335 & 0.3362 & 0.3164 & 0.3128 \\
    \rowcolor{modelgray} Qwen3-8B      & 12 & \textbf{0.3591} & 0.3405 & 0.3311 & 0.3295 & 0.3296 \\
    \rowcolor{modelgray} Qwen3-8B      & 48 & 1.647 & 0.3503 & 0.3250 & 0.3209 & 0.3204 \\
    \rowcolor{modelgray} Qwen3-8B      & 192 & -- & 0.6794 & 0.3386 & \textbf{0.3153} & \textbf{0.3120} \\
    Nemotron-8B   & 12 & 0.3795 & 0.3435 & 0.3318 & 0.3298 & 0.3298 \\
    Nemotron-8B   & 48 & -- & 0.3905 & 0.3336 & 0.3281 & 0.3275 \\
    Nemotron-8B   & 192 & -- & -- & 0.3446 & 0.3174 & 0.3138 \\
    \bottomrule
  \end{tabular}
\end{table}

Two numerical contrasts in Table~\ref{tab:baseline-comparison} are relevant for interpretation.  At $n=500$, the raw-feature GLM has deviance 1.2079, while the best compact embedding GLM has deviance 0.3591; this is consistent with compact embeddings acting as a regularized representation when the design matrix is poorly supported by data.  At $n=500{,}000$, the ranking reverses toward richer representations: Qwen3-8B with $K=192$ reaches 0.3120, compared with 0.3211 for the raw GLM and 0.3194 for Gemini with $K=12$.  Thus compact embeddings perform best in the reported low-label comparisons, whereas large-sample pricing benefits from embedding coordinates that carry more detailed risk variation.

Off-the-shelf embedding choice depends on the available label budget.  In the test results, compact Gemini embedding GLMs are the most stable option at small training sizes.  At the full portfolio size, higher-dimensional Qwen or Nemotron embeddings perform better, with Qwen3-8B at $K=192$ giving the lowest off-the-shelf deviance in Table~\ref{tab:baseline-comparison}.  The raw GLM remains a strong and transparent default, while the embedding GLM provides better predictive performance once it has enough observations to estimate the representation coordinates reliably.

\subsection{Effect of embedding dimensionality}

We investigate the effect of embedding dimensionality on downstream model performance at $n = 500{,}000$.  For each embedding model, we compare two ways to obtain a $K$-dimensional feature vector.  The first applies PCA to the downstream training embedding matrix and keeps the first $K$ principal components.  The second uses the first $K$ embedding coordinates directly, avoiding the PCA step.  For Qwen3-Embedding-0.6B this is the model's Matryoshka prefix; for the other embedding models we treat the same operation more cautiously as a direct-prefix truncation.

\begin{figure}[ht]
  \centering
  \includegraphics[width=\linewidth]{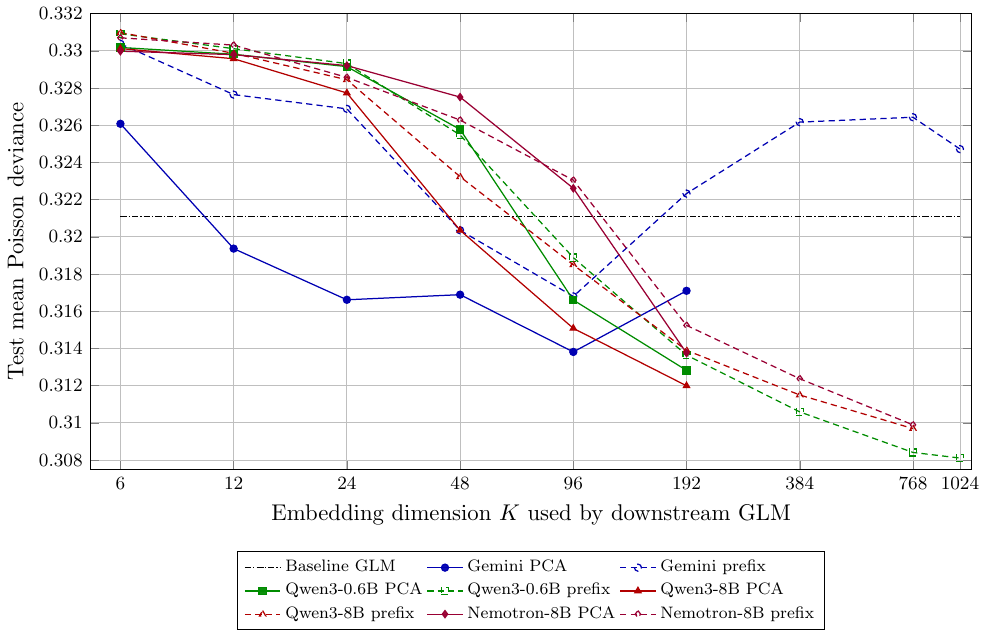}
  \caption{Effect of embedding dimension and reduction method on test deviance across embedding models ($n = 500{,}000$).  Solid curves use PCA components fitted on the downstream training embedding matrix and are available through $K=192$.  Dashed curves use the leading embedding coordinates directly; for Qwen3-0.6B this is the longer Matryoshka-prefix sweep from the single-model experiment, while the other prefix curves stop at their largest evaluated all-model fit.  The horizontal line marks the baseline GLM.}
  \label{fig:reduction-methods-all-models}
\end{figure}

Figure~\ref{fig:reduction-methods-all-models} shows that the PCA and direct-prefix curves answer different questions.  On the paired grid $K \in \{6,12,24,48,96,192\}$ (we stop at 192 since we run into memory issues for larger $K$), PCA is usually at least as strong as the direct prefix because the projection directions are fitted to variation among our specific prompt format.  Exceptions are few; Qwen3-0.6B is slightly better as a prefix at $K=48$, and Nemotron is slightly better as a prefix at $K=24$ and $K=48$.  Gemini illustrates the risk of using a large prefix: its direct-prefix curve improves through $K=96$ and then starts to overfit, even though the training fit continues to improve.  Qwen3-8B and Nemotron benefit from larger direct prefixes through $K=768$, while the Qwen3-0.6B Matryoshka-prefix sweep continues to improve through $K=1024$.  

\subsection{Combining raw features with embeddings}

The dimensionality results show that additional embedding coordinates can help when the training set is large enough, but they also increase the risk of unstable fits in the low-data regime.  Another way to enrich the embedding-based model is to reintroduce the original hand-engineered features alongside a smaller set of embedding components.  We therefore compare whether a fixed budget of additional covariates beyond the 48 raw features in the baseline GLM is better spent on PCA components from the LLM embeddings or on additional embedding dimensions in an embedding-only model.

We compare two Qwen-based strategies as a function of embedding dimension at three training sizes.  In the first strategy, we augment the 48 raw features from the baseline GLM with $K$ PCA components from the Qwen embeddings, yielding a GLM with $48 + K$ covariates.  In the second strategy, we use the leading Qwen Matryoshka prefix directly as the covariate vector, without raw features or PCA.  Both strategies are compared against the baseline GLM at the corresponding training size.

\begin{figure}[ht]
  \centering
  \includegraphics[width=\linewidth]{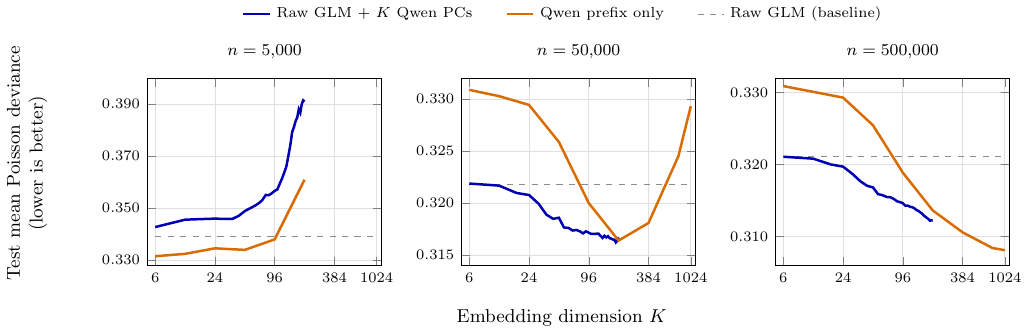}
  \caption{Test mean Poisson deviance for two ways of combining Qwen embeddings with the raw-feature GLM, plotted against the embedding dimension $K$ (log scale) at three downstream training sizes.}
  \label{fig:raw-plus-embeddings}
\end{figure}

Figure~\ref{fig:raw-plus-embeddings} displays the test mean Poisson deviance for the two strategies at each training size, with the raw-feature GLM drawn as a dashed reference line; lower is better.  At $n = 5{,}000$, appending PCA components to the raw GLM hurts, while a small Qwen prefix improves on the raw baseline.  At $n = 50{,}000$, the two strategies are nearly tied at their best dimensions.  At $n = 500{,}000$, both strategies improve over the raw GLM, but the large Qwen prefix gives the lowest deviance.

Raw-plus-embedding models are most useful once the downstream sample is large enough to estimate the extra coefficients.  At $n=5{,}000$, appending embedding PCs to the raw GLM is too unstable, so the better embedding-based option is a compact direct prefix.  The raw-plus-PCs curve rising above the raw-GLM line in Figure~\ref{fig:raw-plus-embeddings} shows that adding PCA components to an already parameterized GLM can increase overfit and worsen test deviance.  At $n=50{,}000$ the two strategies are comparable, suggesting that the raw covariates and embedding PCs contain overlapping signal.  At $n=500{,}000$ a large Qwen prefix gives the lowest deviance.  If auditability of the original variables is central, raw plus a moderate number of embedding PCs is the more interpretable compromise.  If the objective is pure predictive performance at large $n$, the direct Qwen prefix is stronger in these experiments.

\subsection{More flexible downstream models}

The experiments above use a Poisson GLM as the downstream model, which cannot learn non-linear transformations or interactions on its own.  We assess whether the embedding advantage persists when the downstream model is flexible enough to discover these patterns from raw features.  To test this, we replace the GLM with gradient-boosted trees (XGBoost) and compare both downstream models across the 48 engineered raw features and $K = 48$ embedding PCs from Gemini and Qwen.  Hyperparameters are tuned via five-fold cross-validation over a random grid of 12 candidate configurations, using mean validation Poisson deviance; the selected configuration is retrained on the full selected training subset and evaluated once on the test set.

Table~\ref{tab:flexible-models} reports the results of this experiment.  XGBoost improves over the raw-feature GLM at all three training sizes.  For embedding features, the comparison is more mixed: Gemini PCs perform best at $n = 5{,}000$, while Qwen PCs with XGBoost improve substantially at $n = 500{,}000$.  We restrict the table to the matched $K=48$ grid so that each XGBoost row has a corresponding GLM comparison.

\begin{table}[ht]
  \centering
  \caption{Test mean Poisson deviance for GLM and XGBoost downstream models across matched feature sets.  Embedding rows use $K = 48$ PCA components.}
  \label{tab:flexible-models}
  \begin{tabular}{lccc}
    \toprule
    Configuration & $n = 5{,}000$ & $n = 50{,}000$ & $n = 500{,}000$ \\
    \midrule
    \multicolumn{4}{l}{Raw features} \\
    \quad GLM      & 0.3393 & 0.3218 & 0.3211 \\
    \quad XGBoost  & 0.3254 & 0.3188 & 0.3145 \\
    \midrule
    \multicolumn{4}{l}{Gemini embeddings} \\
    \quad GLM      & 0.3226 & 0.3174 & 0.3169 \\
    \quad XGBoost  & 0.3191 & 0.3179 & 0.3269 \\
    \midrule
    \multicolumn{4}{l}{Qwen embeddings} \\
    \quad GLM      & 0.3325 & 0.3265 & 0.3258 \\
    \quad XGBoost  & 0.3304 & 0.3215 & 0.3142 \\
    \bottomrule
  \end{tabular}
\end{table}

The evidence beyond the GLM setting is mixed: embeddings can help a flexible learner in some settings, but raw-feature XGBoost is a strong benchmark, improving on the raw-feature GLM at all three training sizes.  At $K=48$, Gemini embeddings with XGBoost give the best result at $n=5{,}000$ (0.3191), improving on both the raw-feature XGBoost model and the Gemini embedding GLM.  At $n=500{,}000$, however, the same Gemini-XGBoost combination deteriorates to 0.3269, while Qwen embeddings with XGBoost improve to 0.3142 and slightly outperform raw-feature XGBoost.  The large-sample Gemini result suggests that a flexible learner can be sensitive to representation noise or to the selected hyperparameters; it is specific to that model-feature combination.  For an actuarial workflow that values inference and coefficient-level auditability, the embedding GLM remains the cleaner choice; for pure prediction with abundant data, XGBoost on raw or Qwen embedding features is competitive.

\subsection{Sensitivity analyses}

We conduct two sensitivity analyses to assess the robustness of the embedding approach. The first examines differences in prompt format, while the second assesses variable importance through an ablation study. 

\subsubsection{Prompt format sensitivity}\label{sec:prompt-format}

The prompt provided in Figure \ref{fig:prompt-baseline} is just one possible way to encode the policyholder data. Different prompts may improve the quality of the embeddings. We now compare different prompt formats and examine its effect on model performance. Note that, everytime one changes the prompt format, one has to fit a new GLM, since changing the prompt will project the embeddings into a different embedding space. We use Qwen3-Embedding-0.6B with $K = 48$ PCA components and compare six prompt formats applied to the same policyholder data.  We use this model because it is locally deployable and is the model used in the later fine-tuning and ablation diagnostics.  The formats are the default underwriting-list template used in our main experiments (Figure~\ref{fig:prompt-baseline}), a prose narrative, minimal key-value pairs, a task-instruction list, a qualitative descriptor format, and a long underwriting-context format.  The exact text of each prompt template is shown in Appendix~\ref{app:prompts}, along with design rationale.  The qualitative descriptor format is motivated by the observation that LLMs may process qualitative descriptions more effectively than raw numbers: a density of 1,020 people/km$^2$ is tokenized as arbitrary digits, whereas ``high-density urban area'' supplies an explicit qualitative scale that may be easier for the embedding model to encode than the isolated value.  The underwriting-context format tests whether explicitly naming actuarial risk patterns changes the embedding in a way that helps the downstream GLM.  Table~\ref{tab:prompt-format} reports the test deviance for each format across training sizes. 
Training sizes start at $n = 1{,}000$ because several $n = 500$ prompt-format fits were numerically unstable. The broad message is that prompt format matters, but mainly through the way it makes the covariates legible to the embedding model: qualitative descriptors and task instructions are generally the most reliable formats, while the minimal key-value serialization is consistently weaker once the training sample is large enough for the downstream GLM to stabilize.  The long underwriting-context prompt does not dominate the simpler formats, which suggests that the useful part of prompt engineering is to express each policyholder field in a form whose scale and meaning are clear. 

\begin{table}[ht]
  \centering
  \caption{Test mean Poisson deviance by prompt format for Qwen3-Embedding-0.6B across training sizes ($K = 48$). }
  \label{tab:prompt-format}
  \setlength{\tabcolsep}{2.6pt}
  \begin{tabular}{@{}>{\raggedright\arraybackslash}p{1.48in}ccccccccc@{}}
    \toprule
    Prompt format & \multicolumn{9}{c}{Training set size $n$} \\
    \cmidrule(l){2-10}
    & 1{,}000 & 2{,}000 & 5{,}000 & 10{,}000 & 20{,}000 & 50{,}000 & 100{,}000 & 200{,}000 & 500{,}000 \\
    \midrule
    Underwriting list & \textbf{0.4310} & 0.3589 & 0.3325 & 0.3307 & 0.3280 & 0.3265 & 0.3260 & 0.3259 & 0.3258 \\
    Prose narrative          & 0.4398 & 0.3554 & 0.3328 & 0.3293 & 0.3276 & 0.3256 & 0.3253 & 0.3251 & 0.3250 \\
    Minimal key-value        & 0.4721 & 0.3710 & 0.3367 & 0.3334 & 0.3312 & 0.3286 & 0.3282 & 0.3280 & 0.3279 \\
    Task-instruction list & 0.4596 & \textbf{0.3533} & 0.3301 & 0.3276 & 0.3250 & 0.3231 & 0.3236 & 0.3230 & 0.3229 \\
    Qualitative descriptors  & 0.4330 & 0.3613 & \textbf{0.3290} & \textbf{0.3261} & \textbf{0.3239} & \textbf{0.3215} & \textbf{0.3211} & \textbf{0.3206} & \textbf{0.3206} \\
    Underwriting context     & 0.4760 & 0.3570 & 0.3342 & 0.3309 & 0.3284 & 0.3267 & 0.3264 & 0.3265 & 0.3263 \\
    \bottomrule
  \end{tabular}
\end{table}

For subsequent pricing experiments, the results suggest that field-level interpretability is more useful than verbosity.  The qualitative-descriptor format has the lowest reported deviance once the downstream sample reaches $5{,}000$ policies and remains lowest through $n=500{,}000$.  The task-instruction list is the best fallback among these reported results when one wants to keep raw field values but still tell the embedding model the actuarial task.  The minimal key-value format may be attractive when token cost is the main constraint, but in these runs it had worse deviance once the GLM stabilized.  The underwriting-context row is also informative: adding a generic actuarial preamble performs worse than concise field-level descriptions, so prompt engineering should prioritize how each covariate is represented rather than how much domain explanation surrounds it.  These prompt-format differences are small, however, so the method derived in this paper is robust to specific prompt formats. 

\subsubsection{Feature ablation}

We assess how sensitive the embedding-based model is to missing features by dropping one feature at a time from the serialization prompt and re-generating embeddings.  If a model's performance decreases when a feature is removed, this indicates that the feature was important in the model. The ablation uses Qwen3-Embedding-0.6B with the task-instruction list prompt from Table~\ref{tab:prompt-format}.  For each of the nine features, we compute the test deviance of the embedding GLM ($K = 48$) trained on $n = 500{,}000$ rows.  As a reference, we also report the corresponding raw-feature GLM ablation, where the same feature is removed from the engineered baseline design.  Table~\ref{tab:ablation} reports the results.  These ablations are not isolated feature importances, since removing a field changes the whole prompt and therefore the entire embedding geometry. BonusMalus remains the most important field for the raw-feature GLM, while the embedding GLM is especially sensitive to vehicle age and BonusMalus, indicating that these fields drive much of the variation in the embedding representation.  In contrast, removing some geographic fields has little effect, and in some cases slightly improves the embedding model, which is consistent with the strong redundancy among Area, Region, and Density in this dataset.

\begin{table}[ht]
  \centering
  \caption{Feature ablation using Qwen3-Embedding-0.6B and the task-instruction list prompt: test deviance when each feature is dropped from the prompt for the embedding GLM, or from the engineered covariates for the raw-feature GLM. }
  \label{tab:ablation}
  \begin{tabular}{lcc}
    \toprule
    Dropped feature & Embedding GLM & Baseline GLM \\
    \midrule
    None (full model) & 0.3229 & 0.3211 \\
    DrivAge           & 0.3205 & 0.3223 \\
    BonusMalus        & 0.3249 & 0.3276 \\
    VehPower          & 0.3216 & 0.3212 \\
    VehAge            & 0.3263 & 0.3219 \\
    VehBrand          & 0.3219 & 0.3214 \\
    VehGas            & 0.3231 & 0.3211 \\
    Density           & 0.3216 & 0.3211 \\
    Area              & 0.3229 & 0.3211 \\
    Region            & 0.3199 & 0.3212 \\
    \bottomrule
  \end{tabular}
\end{table}

BonusMalus and vehicle age remain important for the embedding model, and BonusMalus remains the most important ablated field for the raw GLM.  The non-monotone entries, such as the improvement after dropping DrivAge or Region from the embedding prompt, should be read as evidence that prompt ablations change the geometry of the whole representation.  They can reduce noise or redundancy as well as remove signal.  The weaker effect of dropping Area, Region, or Density is therefore more consistent with redundancy among geographic encodings than with geographic information being irrelevant.

\section{Fine-tuning the embedding model}\label{s:finetuning-experiments}

We now consider whether fine-tuning the embedding models improves downstream performance. The technical background for LoRA and the CoSENT objective is given in Section~\ref{ss:finetuning}; we test the procedure empirically below. 

\subsection{Procedure}\label{ss:finetuning-procedure}

Off-the-shelf embedding models encode general semantic similarity, but insurance pricing requires that the embedding geometry reflect risk similarity.  We fine-tune the Qwen3-Embedding-0.6B model so that policyholders with similar claim frequencies are mapped to nearby points in the embedding space.

We use Low-Rank Adaptation (LoRA; \citep{hu2022lora}) with the parameterization of \eqref{eq:lora}.  The reference LoRA setting uses rank $r = 8$ and scaling parameter $\alpha = 32$; at this rank, the trainable matrices amount to 0.84\% of the model's total parameters.  These values define the baseline adapter setting.  Table~\ref{tab:finetuning-results} reports additional LoRA configurations, including rank sensitivity rows, so the fine-tuning section should be read as a small configuration grid.  The fine-tuning objective is the CoSENT loss of \eqref{eq:cosent} applied to the pairwise similarity scores $s_{ij}$ of \eqref{eq:similarity}, computed from the observed claim frequencies $f_i = N_i/E_i$ with $E_i > 0$; tied pair labels are ignored in the ordered comparisons.  After fine-tuning, the embedding function $f_{\text{embed}}$ produces vectors whose geometry is aligned with claim frequency patterns in the fine-tuning data.

We record the remaining fine-tuning hyperparameters here for reproducibility.  The similarity temperature in \eqref{eq:similarity} is $\tau = 0.35$, and the scaled CoSENT loss in \eqref{eq:cosent} uses $\lambda = 20$.  LoRA is applied to the attention projection matrices; the feed-forward (MLP) sub-layers are left frozen.  

The CoSENT loss requires data pairs. We uniformly generate such pairs; for each pair, the two members are drawn independently with replacement from the 10,000-policy fine-tuning pool. Because roughly 95\% of the fine-tuning policies have no claims, most sampled pairs join two zero-claim policies and are tied at $s_{ij} = 1$; only about 10\% of the sampled pairs are informative (non-tied) and contribute ordered comparisons to the CoSENT loss.  

A limitation of this objective is that the individual observed claim frequency $f_i = N_i/E_i$ is a noisy proxy for the latent risk: claims are rare, so most policies have $f_i = 0$, even if a policyholder is riskier than the others, and the observed frequency for a single policy carries little information about its true expected frequency.  Because most policies have zero claims, many sampled pairs are tied in $s_{ij}$ and contribute no ordered comparisons; the informative comparisons come from pairs whose observed frequencies differ.

Fine-tuning uses a held-out set of policies that is disjoint from the canonical downstream training and test splits.  For each run, we sample 10,000 fine-tuning policies and construct paired prompt examples with the labels in \eqref{eq:similarity}.  The LoRA adapter is trained with the CoSENT loss wrapped in a Matryoshka loss at dimensions 48, 192, and 1024, so that the adapted embeddings remain useful after dimensionality reduction.  Let $\mathcal{M}=\{48,192,1024\}$ and let $\mathbf{e}_i^{(m)}$ denote the first $m$ coordinates of $\mathbf{e}_i$.  The training objective is the equally weighted average
\[
  \mathcal{L}_{\mathrm{MRL}}
  =
  \frac{1}{|\mathcal{M}|}\sum_{m\in\mathcal{M}}
  \mathcal{L}_{\mathrm{CoSENT}}\!\left(\{\mathbf{e}_i^{(m)}\},\{s_{ij}\};\lambda\right).
\]
The downstream comparison then embeds the canonical train-test prompts with the frozen or adapted Qwen model, fits PCA on the selected downstream training subset, and fits the same Poisson GLM with log exposure offset as in the main experiments.  The fine-tuning set is never used to fit the downstream pricing GLM.

The core LoRA variants built around the reference adapter setting all use Qwen3-Embedding-0.6B, 10,000 fine-tuning policies, and learning rate $10^{-5}$: 50,000 pairs for 1 epoch, 50,000 pairs for 2 epochs, and 100,000 pairs for 1 epoch.  These variants test whether the adapter benefits more from additional epochs or more sampled pairs.  The downstream comparison also includes rank-labeled sensitivity runs, longer 100,000-pair runs, and larger 200,000- and 300,000-pair runs; these are reported below because they clarify which fine-tuning choices matter most empirically.

\subsection{Results}\label{ss:finetuning-results}

\begin{figure}[ht]
  \centering
  \includegraphics{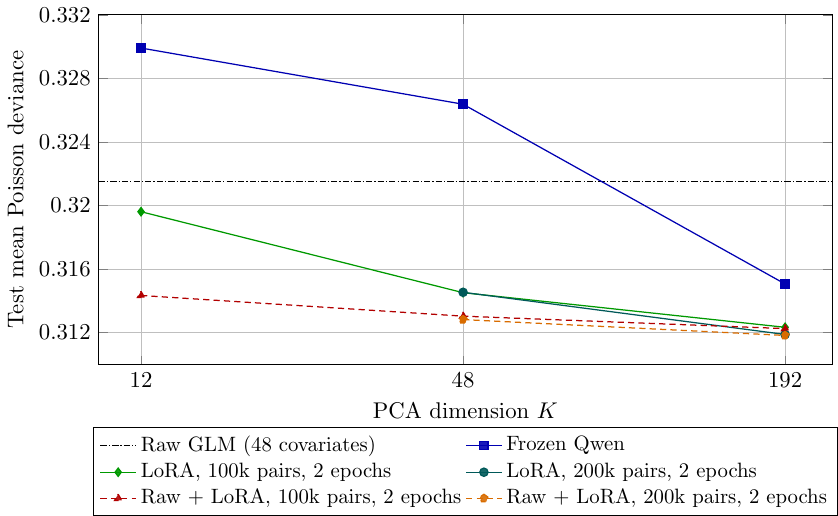}
  \caption{Fine-tuning effect for Qwen3-0.6B at $n=70{,}000$: test mean Poisson deviance as a function of the PCA dimension $K$. The dash-dotted horizontal line marks the raw-feature GLM.}
  \label{fig:finetuning-curves}
\end{figure}

Figure~\ref{fig:finetuning-curves} shows the main effect of fine-tuning for the two strongest adapters in the evaluated grid.  Fine-tuning shifts the Qwen PCA curve downward at every tested dimension, with the largest gain at $K=48$.  The 100,000-pair, two-epoch adapter provides the strongest low-dimensional embedding-only result, while the 200,000-pair, two-epoch adapter gives the best high-dimensional result.  Adding the raw GLM covariates to the fine-tuned PCs gives another improvement, especially at the smaller dimensions where the embedding representation has fewer coordinates to carry the structured actuarial signal.  At $K=192$, the gap between embedding-only and raw-plus-embedding becomes small, suggesting that the adapted high-dimensional embedding carries much of the information contained in the engineered covariates.

Table~\ref{tab:finetuning-results} reports the downstream test deviance for the frozen Qwen embedding, the fine-tuned variants, and combinations of the raw GLM covariates with fine-tuned embedding PCs.  These comparisons use $n=70{,}000$ downstream training policies and the full 100,000-policy test split.  Fine-tuning improves over frozen Qwen at every evaluated matched dimension.  Among the reported embedding-only test results, the 100,000-pair and 200,000-pair two-epoch adapters are tied at $K=12$ to the displayed precision, the 100,000-pair two-epoch adapter is lowest at $K=48$, and the 200,000-pair two-epoch adapter is best at $K=192$.  When raw covariates are appended, the 200,000-pair two-epoch adapter has the lowest reported deviance at both $K=48$ and $K=192$, with the best overall $n=70{,}000$ result equal to 0.3118 (raw features plus fine-tuned embedding PCA at $K=192$).

\begin{table}[ht]
  \centering
  \caption{Fine-tuned Qwen downstream comparison: test mean Poisson deviance at $n=70{,}000$.  Each LoRA adapter is listed by its number of training pairs, epoch count, and rank; the rank column records the configuration used for that adapter, including the rank-sensitivity rows.  Dashes indicate not-applicable entries.  Bold marks the lowest displayed deviance in each column within each covariate block.}
  \label{tab:finetuning-results}
  \begin{tabular}{@{}lccccccc@{}}
    \toprule
    Model & Pairs & Epochs & Rank & $K=12$ & $K=48$ & $K=192$ \\
    \midrule
    \multicolumn{7}{@{}l}{\textit{Raw features}} \\
    \quad Raw GLM & -- & -- & -- & -- & 0.3215 & -- \\
    \midrule
    \multicolumn{7}{@{}l}{\textit{Embedding PCA}} \\
    \quad Frozen Qwen     & --   & -- & --  & 0.3299 & 0.3264 & 0.3151 \\
    \quad LoRA            & 50k  & 1  & 8   & 0.3245 & 0.3180 & 0.3138 \\
    \quad LoRA            & 50k  & 2  & 8   & 0.3222 & 0.3161 & 0.3130 \\
    \quad LoRA            & 50k  & 2  & 32  & 0.3210 & 0.3156 & 0.3128 \\
    \quad LoRA            & 100k & 1  & 8   & 0.3207 & 0.3152 & 0.3124 \\
    \quad LoRA            & 100k & 2  & 8   & \textbf{0.3196} & \textbf{0.3145} & 0.3123 \\
    \quad LoRA            & 100k & 3  & 8   & 0.3211 & 0.3152 & 0.3125 \\
    \quad LoRA            & 100k & 4  & 8   & 0.3228 & 0.3166 & 0.3131 \\
    \quad LoRA            & 200k & 2  & 8   & \textbf{0.3196} & 0.3145 & \textbf{0.3119} \\
    \quad LoRA            & 300k & 2  & 8   & 0.3206 & 0.3154 & 0.3124 \\
    \midrule
    \multicolumn{7}{@{}l}{\textit{Raw + embedding PCA}} \\
    \quad LoRA            & 50k  & 1  & 8   & 0.3168 & 0.3147 & 0.3136 \\
    \quad LoRA            & 50k  & 2  & 8   & 0.3148 & 0.3132 & 0.3129 \\
    \quad LoRA            & 50k  & 2  & 32  & \textbf{0.3141} & 0.3133 & 0.3129 \\
    \quad LoRA            & 100k & 1  & 8   & 0.3145 & 0.3137 & 0.3126 \\
    \quad LoRA            & 100k & 2  & 8   & 0.3143 & 0.3130 & 0.3123 \\
    \quad LoRA            & 100k & 3  & 8   & 0.3149 & 0.3136 & 0.3124 \\
    \quad LoRA            & 200k & 2  & 8   & 0.3145 & \textbf{0.3128} & \textbf{0.3118} \\
    \quad LoRA            & 300k & 2  & 8   & 0.3143 & 0.3131 & 0.3123 \\
    \bottomrule
  \end{tabular}
\end{table}

Table \ref{tab:finetuning-results} also clarifies how adaptation changes the dimension tradeoff.  The largest relative gains occur at smaller dimensions: at $K=48$, the best embedding-only adapter reduces frozen Qwen from 0.3264 to 0.3145, slightly better than the frozen $K=192$ result of 0.3151.  Raw covariates help most when the embedding is compressed, reducing the best $K=12$ adapter from 0.3196 to 0.3143, but they add little once the adapted embedding reaches $K=192$; for the 200,000-pair two-epoch adapter, the embedding-only and raw-plus-embedding deviances are 0.3119 and 0.3118, respectively.  Increasing the pair count from 100,000 to 200,000 improves the high-dimensional representation, but the gains are not monotone: the 300,000-pair two-epoch run is worse than the 200,000-pair run at all three reported dimensions, and the 100,000-pair three- and four-epoch rows do not improve on the 100,000-pair two-epoch row.  This suggests that fine-tuning improves the risk relevance of the leading embedding components, but longer or larger adaptation can overfit the held-out fine-tuning signal. 

\begin{figure}[ht]
  \centering
  \includegraphics{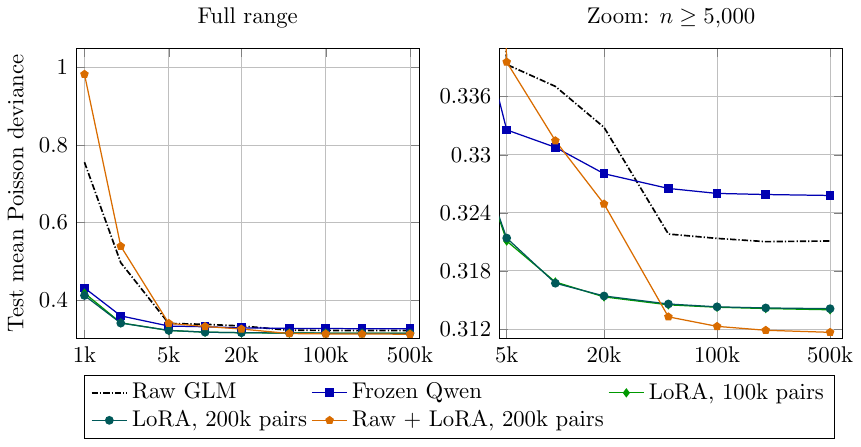}
  \caption{Learning curves for Qwen fine-tuning at $K=48$.  The left panel shows the full range from $n=1{,}000$ to $n=500{,}000$; the right panel zooms in on the stable medium- and large-sample regime.  Raw-plus-LoRA appends the 48 GLM covariates to the 48 fine-tuned PCA components.  The 100,000-pair and 200,000-pair adapters shown are the two-epoch variants, matching the labelling of Table~\ref{tab:finetuning-results}.}
  \label{fig:finetuning-learning-curves}
\end{figure}

The comparison also records learning curves for the main fine-tuned adapters at $K=48$.  Figure~\ref{fig:finetuning-learning-curves} shows that fine-tuning helps the embedding-only Qwen GLM across the downstream training sizes where the fit is stable: both the 100,000-pair and 200,000-pair adapters improve substantially over frozen Qwen from $n=1{,}000$ onward.  The two fine-tuned embedding-only curves are nearly tied at $K=48$, with the 200,000-pair adapter slightly better at $n=1{,}000$, $2{,}000$, and $10{,}000$, and the 100,000-pair adapter slightly better at $n\geq 20{,}000$.  Appending the raw GLM covariates to the fine-tuned PCs is harmful at small and moderate sample sizes, but becomes the best $K=48$ specification once the $n = 50,000$.  This is the same variance tradeoff seen elsewhere in our experiments: raw covariates add useful actuarial signal when the downstream GLM has enough observations, but they create too many coefficients when the training subset is small.

Table~\ref{tab:finetuning-large-sample} examines whether the fine-tuning gain persists when the downstream GLM is trained on 500,000 policies.  The answer is yes: at $K=192$, the 100,000-pair two-epoch adapter reduces the frozen Qwen test deviance from 0.3128 to 0.3098, and appending raw covariates reduces it further to 0.3094.  The 200,000-pair rows at $K=48$ do not improve the embedding-only large-sample result, but they slightly improve the raw-plus-embedding result from 0.3118 to 0.3117.  Thus fine-tuning changes the representation in a way that remains useful in the large-sample regime, while the best full-data specification in this comparison is the $K=192$ raw-plus adapter; the 200,000-pair adapter was not part of the $K=192$ full-data comparison.

Among the reported $n=70{,}000$ test results, the 200,000-pair, two-epoch Qwen LoRA adapter with $K=192$ PCA components is strongest, optionally augmented with the raw GLM covariates when coefficient-level access to the original variables is useful.  The margins separating the strongest adapters are nonetheless small.  In the reported runs, increasing the LoRA rank is less effective than increasing the number of informative pairs, but the 300,000-pair and longer-epoch rows show that additional adaptation has diminishing returns.  In these experiments, fine-tuning was the most effective way to make the local Qwen model competitive with stronger frozen off-the-shelf alternatives while retaining local deployment and adaptation.

Because fine-tuning consumes the claim outcomes of 10,000 held-out policies, the fine-tuned pipeline evaluated at $n = 70{,}000$ uses an effective budget of 80,000 labeled policies.  To check whether the improvement is merely an effect of these extra labels, Table~\ref{tab:label-budget} gives the label-matched comparators: the frozen-Qwen GLM and the raw-feature GLM trained on $n = 80{,}000$ policies, evaluated at $K = 192$.  The fine-tuned adapter trained on only 70,000 downstream policies still beats the frozen model at the matched 80,000-policy budget and the raw GLM at 80,000.  The 70,000-to-80,000 fine-tuned gap is itself small, so the improvement over the frozen and raw models is not explained by the 10,000 additional labels.

\begin{table}[ht]
  \centering
  \caption{Label-budget-matched comparison at $K = 192$: test mean Poisson deviance.  The fine-tuned 100,000-pair, two-epoch adapter at $n = 70{,}000$ (effective budget 80,000) is compared against frozen and raw baselines given the full 80,000-policy budget.}
  \label{tab:label-budget}
  \begin{tabular}{llc}
    \toprule
    Model & $n$ (downstream) & $K = 192$ \\
    \midrule
    Fine-tuned (100k-pair, 2 epochs) & 70{,}000 & 0.3123 \\
    Fine-tuned (100k-pair, 2 epochs) & 80{,}000 & 0.3117 \\
    Frozen Qwen                      & 80{,}000 & 0.3147 \\
    Raw GLM                          & 80{,}000 & 0.3214 \\
    \bottomrule
  \end{tabular}
\end{table}

\begin{table}[ht]
  \centering
  \caption{Large-sample fine-tuned Qwen comparison: test mean Poisson deviance at $n=500{,}000$.  Bold marks the lowest deviance in each column.}
  \label{tab:finetuning-large-sample}
  \begin{tabular}{llcc}
    \toprule
    Covariates & Embedding or model & $K=48$ & $K=192$ \\
    \midrule
    Raw features & Raw GLM & 0.3211 & -- \\
    Embedding PCA & Frozen Qwen & 0.3258 & 0.3128 \\
    Embedding PCA & LoRA, 100k pairs, 1 epoch & 0.3146 & 0.3100 \\
    Raw + embedding PCA & LoRA, 100k pairs, 1 epoch & 0.3123 & 0.3097 \\
    Embedding PCA & LoRA, 100k pairs, 2 epochs & 0.3140 & 0.3098 \\
    Embedding PCA & LoRA, 200k pairs, 2 epochs & 0.3141 & -- \\
    Raw + embedding PCA & LoRA, 100k pairs, 2 epochs & 0.3118 & \textbf{0.3094} \\
    Raw + embedding PCA & LoRA, 200k pairs, 2 epochs & \textbf{0.3117} & -- \\
    \bottomrule
  \end{tabular}
\end{table}

\section{Discussion and conclusion}\label{s:discussion}

This paper studied whether LLM embeddings of serialized policyholder descriptions can serve as covariates in a conventional Poisson GLM for claim frequency pricing.  On the French MTPL data, embedding-based GLMs can improve on the baseline GLM, with the largest gains in the low-data regime.  Ranking-based contrastive fine-tuning with LoRA improves the locally deployable Qwen embeddings at every tested PCA dimension.  Prompt design matters: formats that make the scale and meaning of each covariate semantically relevant to the embedding model outperform both minimal serializations and long generic preambles.  

A central feature of the pipeline is that it is deterministic.  The LLM receives no claims data at inference time and generates no text; it maps each policyholder description to a fixed embedding, and the premium is set entirely by the GLM.  There is no sampling, no stochastic decoding, and hence no possibility of hallucination: the same description always yields the same covariates and the same price.  This distinguishes the approach from generative applications that emit predictions as free text and simplifies governance, since embeddings can be computed once per policyholder, and treated as a fixed preprocessing step upstream of a GLM that is fitted by maximum likelihood and audited with existing regulatory tools.  

The main limitations concern interpretability and bias.  The downstream estimation is transparent, but the PCA components derived from the embeddings are not directly interpretable in terms of policyholder characteristics. Models combining the original variables with a moderate number of embedding components offer a governance compromise, since the original covariates retain interpretable coefficients.  The embeddings also inherit biases present in the LLM's pre-training data: the protected-attribute diagnostic of Appendix~\ref{s:fairness} shows that the pipeline reacts to text appended to the prompt, with neutral out-of-template fields shifting predictions by at least as much as demographic tokens.  Controlled prompt templates, protected-attribute exclusion rules, and perturbation audits are therefore a governance requirement.

As an effort to interpret the embedding, we performed a clustering analysis of the embedding vectors of the textual claims in Appendix~\ref{app:clustering}. The clusters provided pricing bands spanning $0.79\times$ to $1.52\times$ the portfolio claim frequency and recovers a brand, area, and vehicle-age segmentation, so the embedding carries risk-relevant structure on its own, although the supervised GLM exploits it more fully.

\section*{Acknowledgements}

CBW acknowledges financial support from the Natural Sciences and Engineering Research Council of Canada (RGPIN-2025-06879). The project is also funded by the Society of Actuaries Research Institute -- University AI Research Projects.





\bibliographystyle{apalike}
\bibliography{ref}

\appendix

\section{Prompt-invariance diagnostic}\label{s:fairness}

A concern with using LLM embeddings for pricing is that the model's pre-trained knowledge may encode societal biases, causing protected attributes to influence premiums even when they are not legitimate rating factors.  We test this sensitivity by augmenting the prompt for each of 100,000 test policies while holding the downstream pricing model fixed.  For each baseline prompt, we generate five perturbed prompts: two appending a race token (``Race: White'' or ``Race: Black''), two appending a neutral out-of-template control of comparable length (``Favourite colour: blue'' or ``Policyholder height: 175 cm''), and one increasing the displayed BonusMalus value by 10 points. The BonusMalus perturbation is included as a positive control, since it changes a risk-relevant pricing covariate; the race and neutral-control perturbations should ideally leave the prediction unchanged.

\begin{figure}[ht]
  \centering
  \includegraphics[width=\linewidth]{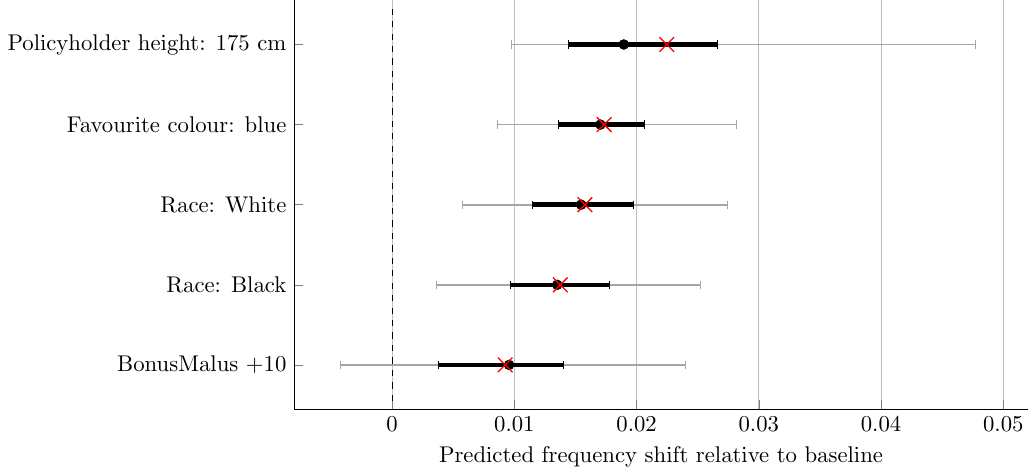}
  \caption{Distribution summaries of individual-level predicted frequency shifts for the five prompt perturbations ($n=100{,}000$ test policies): two race tokens, two neutral out-of-template controls (favourite colour, policyholder height), and the BonusMalus positive control.  Points show medians, thick intervals show interquartile ranges, thin intervals show 5th--95th percentile ranges, and red crosses show means.}
  \label{fig:fairness-perturbations}
\end{figure}

This is not a demographic fairness study; it is a prompt-invariance stress test of how sensitive the embedding pipeline is to tokens that should not affect pricing.  The same policyholders are re-scored under synthetic prompt additions, so the results do not estimate claim frequency by race or measure the demographic fairness of premiums.  The race perturbations are deliberately out-of-distribution relative to the controlled production prompt, since the diagnostic probes what happens when the template is not controlled.  The standardization, PCA centering and loading matrix, and GLM coefficients are fitted once on the original downstream training embeddings and held fixed for all perturbed test embeddings: the fixed pipeline is the frozen Qwen3-Embedding-0.6B model with the default underwriting-list prompt of Figure~\ref{fig:prompt-baseline}, $K = 48$ PCA components, and a Poisson GLM with log-exposure offset fitted on the $n = 500{,}000$ downstream training pool.  The perturbations are applied to the embeddings of the full 100,000-policy test set, and we compare perturbed and baseline predictions policy by policy, reporting the average shift in predicted frequency and the fraction of policies whose predicted frequency increases, see Figure \ref{fig:fairness-perturbations}.

%

The diagnostic shows that the frozen Qwen representation is sensitive to the synthetic race field.  Adding either race label increases the mean predicted claim frequency, and the average race-induced shifts are larger than the shift from increasing BonusMalus by 10 points.  That sensitivity would need to be ruled out before using unrestricted prompts in pricing.  Both race labels shift predictions in the same direction, and the ``Race: White'' token shifts them more than the ``Race: Black'' token.  This pattern is at least as consistent with the pipeline reacting to any unseen field appended to a fixed template as with race semantics specifically, and the neutral-control perturbations (favourite colour and policyholder height) support this appended-field explanation.

To distinguish race-token sensitivity from sensitivity to any appended out-of-template field, we also include two semantically irrelevant control perturbations of comparable token length: ``Favourite colour: blue'' and ``Policyholder height: 175 cm'', appended to the base prompt in the same position and format as the race tokens and evaluated on the same test policies.  The neutral controls move predictions by at least as much as the race tokens: appending ``Favourite colour: blue'' raises the mean predicted frequency by $16.1\%$ and ``Policyholder height: 175 cm'' by $20.9\%$, both exceeding the $14.6\%$ and $12.8\%$ shifts from the race labels.  Appending any unseen out-of-template field therefore raises predicted frequency by a similar amount, so this probe measures generic sensitivity to appended novel fields, a distribution-shift artifact, rather than race-specific encoding.  Accordingly, we do not interpret the diagnostic as evidence of race-specific behaviour, and we make no race-specific governance claim about the pipeline on this basis. 

These shifts are small in absolute terms, at most about two percentage points of predicted frequency, and race does not stand out, since the race tokens move predictions no more than the semantically irrelevant controls.  Because every perturbation is out-of-template, the diagnostic should be read as a measure of prompt sensitivity rather than as race-specific evidence.

Prompt construction is therefore part of model governance: the same mechanism that lets embeddings absorb new fields also lets them absorb fields that should have no pricing effect, so deployment should use controlled prompt templates, protected-attribute exclusion rules, and perturbation audits before using these methods for insurance pricing.

\section{Unsupervised rating classes from embedding clusters}\label{app:clustering}

The main experiments use the policyholder embedding in a supervised pricing model.  We reduce the embedding by PCA and fit a Poisson GLM against claims.  In this appendix we run a simpler experiment: we cluster policies using only the embeddings generated from the textual prompts.  After the clusters are fixed, we compute the claim frequency in each cluster.  Because the clustering itself never sees claim counts, different frequencies across clusters would mean the embedding already contains risk information, before the supervised GLM is fitted.  We use the same 600,000-policy portfolio as in the main experiments.

\begin{figure}[ht]
  \centering
  \includegraphics[width=\linewidth]{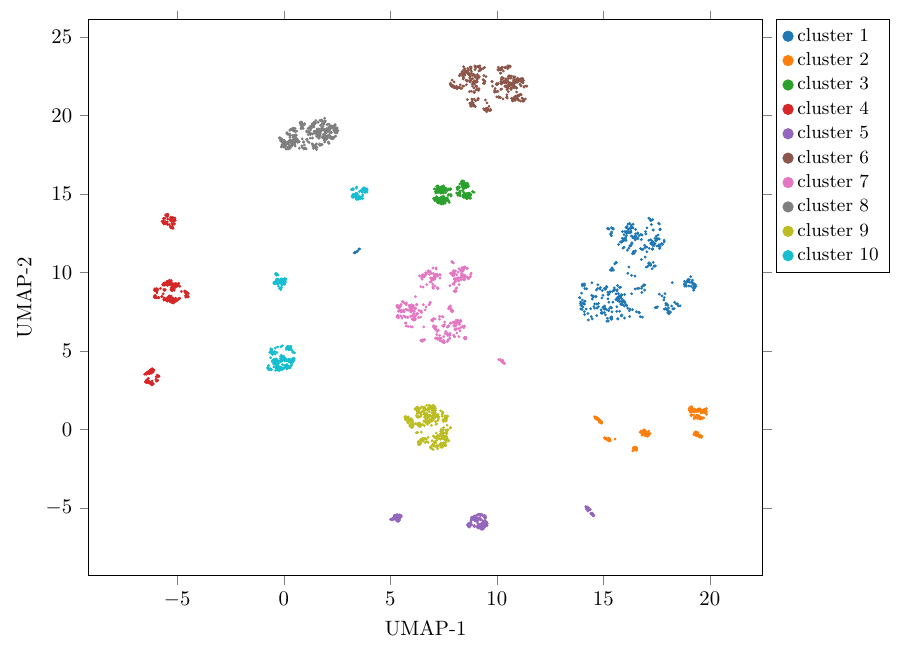}
  \caption{UMAP projection of the Qwen3-Embedding-0.6B policyholder embeddings, colored by $K$-means cluster.}
  \label{fig:umap-clusters}
\end{figure}

We build the clusters in two steps.  First, we serialize each policy with the default underwriting prompt of Section~\ref{s:methodology} and embed it with Qwen3-Embedding-0.6B into a $1024$-dimensional vector.  We take a fixed random sample of $60{,}000$ policies, reduce them to $50$ PCA dimensions ($96.1\%$ of variance retained), and run $K$-means with $K=10$.  These ten clusters are the potential pricing bands.  Second, we assign all $600{,}000$ policies to the nearest centroid in the same $50$-dimensional PCA space.  We then compute each band's claim frequency on the full portfolio.  Figure~\ref{fig:umap-clusters} shows the clusters in two dimensions (a UMAP projection of the 50 principal components), where the groups appear visually separated.  In the full $50$-dimensional space the $K$-means silhouette is only $0.17$; the silhouette runs from $-1$ to $1$, with higher values meaning cleaner separation, so $0.17$ indicates clusters that overlap substantially despite the visual gaps in the projection.  

\begin{figure}[ht]
  \centering
  \includegraphics[width=0.82\linewidth]{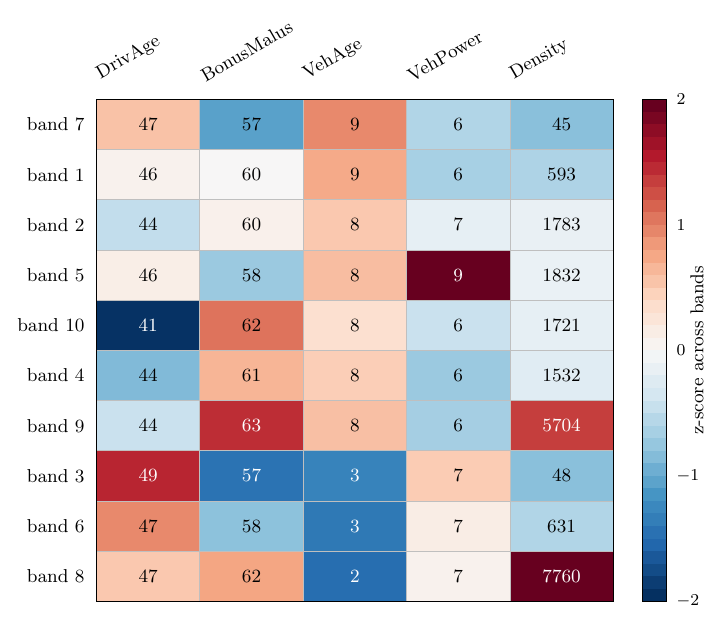}
  \caption{Per-band feature profile.  Each cell prints the mean of that feature over the policies in the band.  The colour shows whether that band is above or below the average band for the same feature, using the $z$-score in \eqref{eq:heatmap-z}.  Bands are ordered from lowest to highest claim frequency.}
  \label{fig:cluster-heatmap}
\end{figure}

Figure~\ref{fig:cluster-heatmap} compares the bands with the original covariates.  We find that most bands are separated mainly by vehicle brand and geography.  Eight of the ten bands are dominated by a single vehicle-brand group, and the other two bands still have one brand group as the largest component.  The bands also differ by area type and population density.  Fuel type does not separate the bands: the diesel share stays between $40\%$ and $60\%$ in every band, so no band is distinctively diesel or gasoline. The heatmap uses colour to show which band means are high or low relative to the other bands.  Each row is a band and each column is a feature.  The number printed in each cell is the mean of that column's feature over the policies in the band.  The colour is a column-wise $z$-score, so variables on different scales can be shown in one figure.  Let $\bar{x}_{bj}$ be the mean of feature $j$ over the policies assigned to band $b$, which is the raw value printed in cell $(b,j)$.  For each feature $j$ we standardize these ten band means across the bands,
\begin{equation}\label{eq:heatmap-z}
  z_{bj} = \frac{\bar{x}_{bj} - \mu_j}{\sigma_j},
  \qquad
  \mu_j = \frac{1}{10}\sum_{b=1}^{10} \bar{x}_{bj},
  \qquad
  \sigma_j = \sqrt{\frac{1}{10}\sum_{b=1}^{10}\bigl(\bar{x}_{bj}-\mu_j\bigr)^2},
\end{equation}
so that $z_{bj}$ measures how far band $b$ is from the average band for feature $j$.  Red means the band mean is above the average for that feature, blue means it is below, and pale means it is close to average.  The printed number remains the raw band mean $\bar{x}_{bj}$.  The colours compare bands with one another; they do not compare a band with the full portfolio.

The clusters mostly reflect vehicle brand, area, and vehicle age.  The three highest-frequency bands (3, 6, 8) are new-car bands, with mean vehicle age $2.4$--$2.9$ years.  The other bands have mean vehicle age $7.5$--$9.3$ years.  We find that density also matters: bands 8 and 9 are urban, while band 7 is rural.  Band 3 is higher-frequency even though it is rural because it is also a new-car band.  Driver age and Bonus-Malus do not separate the bands.  Mean driver age ranges only from $41$ to $49$, Bonus-Malus means range from $57$ to $63$, and the median Bonus-Malus is $50$, the best possible score, almost everywhere.  Thus the clustering mainly recovers a brand $\times$ area $\times$ vehicle-age grouping.  It misses driver age and Bonus-Malus, which are important in classical pricing models.

\begin{figure}[ht]
  \centering
  \includegraphics[width=\linewidth]{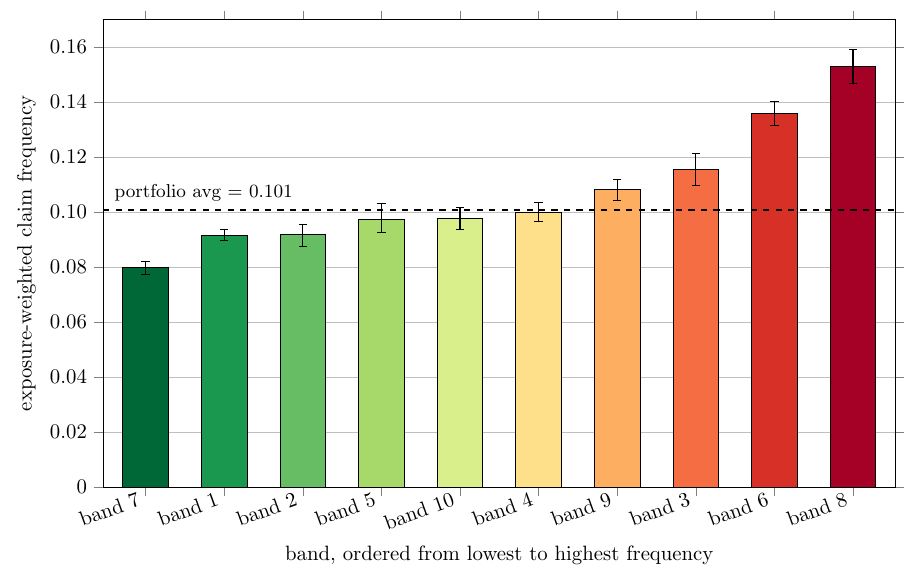}
  \caption{Exposure-weighted claim frequency by band, computed on the full $600{,}000$-policy portfolio and ordered from lowest to highest frequency, with $95\%$ Poisson bootstrap intervals.  The dashed line marks the portfolio average ($0.101$).  The lowest and highest bands are separated, but the six middle bands overlap.}
  \label{fig:rating-bands}
\end{figure}

Figure~\ref{fig:rating-bands} and Table~\ref{tab:rating-bands} order the bands by their exposure-weighted claim frequency.  The lowest band has frequency $0.080$, and the highest has frequency $0.153$.  The highest rate is therefore about $1.9$ times the lowest.  To check stability, we hold the cluster assignments fixed and recompute each band's claim frequency on five disjoint folds of the portfolio; the low and high ends are stable across folds, with band 7 remaining low and bands 6 and 8 remaining high.  The middle bands (1, 2, 5, 10, 4, 9) are close to one another.  Their confidence intervals overlap around the portfolio average, so they should not be read as separate rate levels.  In practice, the clustering gives fewer than ten groups: band 7 is low, the six middle bands are similar, band 3 is somewhat higher, and bands 6 and 8 are high.

\begin{table}[ht]
  \centering
  \caption{Rating bands ordered by full-portfolio exposure-weighted claim frequency, with $95\%$ Poisson bootstrap intervals and the ratio to the portfolio average ($0.1006$).  The band column gives the original $K$-means cluster label.}
  \label{tab:rating-bands}
  \begin{tabular}{ccrccc}
    \toprule
    Frequency rank & Band & $n$ & Claim frequency & $95\%$ CI & vs.\ portfolio \\
    \midrule
    1 (lowest)    & 7  & 88{,}000  & 0.0798 & $[0.077, 0.082]$ & $0.79\times$ \\
    2             & 1  & 144{,}000 & 0.0916 & $[0.090, 0.094]$ & $0.91\times$ \\
    3             & 2  & 40{,}000  & 0.0917 & $[0.087, 0.096]$ & $0.91\times$ \\
    4             & 5  & 28{,}000  & 0.0973 & $[0.093, 0.103]$ & $0.97\times$ \\
    5             & 10 & 47{,}000  & 0.0978 & $[0.094, 0.102]$ & $0.97\times$ \\
    6             & 4  & 53{,}000  & 0.1000 & $[0.096, 0.104]$ & $0.99\times$ \\
    7             & 9  & 54{,}000  & 0.1081 & $[0.104, 0.112]$ & $1.07\times$ \\
    8             & 3  & 31{,}000  & 0.1155 & $[0.110, 0.121]$ & $1.15\times$ \\
    9             & 6  & 71{,}000  & 0.1357 & $[0.131, 0.140]$ & $1.35\times$ \\
    10 (highest)  & 8  & 45{,}000  & 0.1529 & $[0.147, 0.159]$ & $1.52\times$ \\
    \bottomrule
  \end{tabular}
\end{table}

We compare the lowest and highest bands to interpret the policyholder profile.  Band 7 and band 8 have nearly the same driver profile: mean driver age is $47$ in both bands, and median Bonus-Malus is $50$ in both bands.  The difference is the car and the place.  Band 7 contains older Renault, Nissan, or Citro\"en vehicles, with mean vehicle age $9.3$ years, in a low-density rural setting.  Its median density is about $42$ people/km$^2$, and many policies are in the Centre region.  Band 8 contains nearly new Japanese or Korean imports, with mean vehicle age $2.4$ years, in a dense urban setting.  Its median density is about $4{,}350$ people/km$^2$, and about half of the band is in \^Ile-de-France.  It shows that two groups with similar driver age and Bonus-Malus can have very different frequencies when vehicle age, brand, and location differ.

We conclude that the embedding contains risk information before any supervised training.  The clustering never uses claims, but the final bands still range from $0.79\times$ to $1.52\times$ the portfolio frequency.  The low and high bands are also stable across the five folds.  These bands can serve as an initial grouping for a rating book, but they are not a finished pricing model.  What they recover is largely a brand $\times$ area $\times$ vehicle-age grouping already implicit in the structured data, and they leave driver age and Bonus-Malus unseparated.  Because the middle bands' confidence intervals overlap, only a handful of distinct frequency groups survive, not ten rate levels.  The clustering thus serves as a check on the embedding, while the supervised GLM of the main text does the pricing.  

\section{Prompt templates}\label{app:prompts}

The sensitivity analysis in Section~\ref{sec:prompt-format} compares six prompt formats for serializing policyholder covariates into text: the default instruction-prefixed underwriting list, displayed in Figure~\ref{fig:prompt-baseline} of the main text, and the five alternatives displayed below (prose narrative, minimal key-value, task-instruction list, qualitative descriptors, and underwriting context).  Each template is shown using a representative policyholder (DrivAge = 18, Area = E, Region = R24, Density = 1020, VehBrand = B2, VehAge = 1, VehGas = Diesel, VehPower = 7, BonusMalus = 100), with the design rationale for each.  The density anchor of 1,792 people/km$^2$ used in the default prompt is the population-weighted average of the Density field over the processed insurance policy records, not France's national population density.

\subsection{Prose narrative}\label{app:prompt-prose}

\begin{figure}[ht]
  \centering
  \begin{prompt}
An 18-year-old driver living in an urban area, a densely populated
zone in the Centre region with 1020 people per square kilometer.
The driver operates a Renault, Nissan, or Citroën vehicle that is 1 year old,
powered by diesel fuel, in power class 7. The driver's
Bonus-Malus score is 100, indicating the entrance level
with no prior claims history.
\end{prompt}
\caption{Prose narrative prompt.}
\label{fig:prompt-prose}
\end{figure}

The prose format (Figure~\ref{fig:prompt-prose}) embeds the same raw values into natural-language sentences that provide some contextual framing (e.g., ``indicating the entrance level with no prior claims history'' for a BonusMalus of 100).  The hypothesis is that prose changes the token order and adjacency seen by the embedding model, placing related fields such as ``18-year-old driver'' and ``densely populated zone'' close together in a single sentence.

\subsection{Minimal key-value}\label{app:prompt-kv}

\begin{figure}[ht]
  \centering
  \begin{prompt}
DrivAge=18, Area=Urban area, Region=Centre, Density=1020,
VehBrand=Renault/Nissan/Citroën, VehAge=1, VehGas=Diesel, VehPower=7,
BonusMalus=100
\end{prompt}
\caption{Minimal key-value prompt.}
\label{fig:prompt-kv}
\end{figure}

The minimal format (Figure~\ref{fig:prompt-kv}) reduces the prompt to its informational core, using variable names directly from the dataset separated by commas.  This tests whether the verbosity of the prompt matters for embedding quality, or whether the LLM extracts the same semantic information regardless of formatting.  If so, the key-value format would be preferable, since shorter token sequences reduce embedding computation time proportionally.  One concern with this format is that LLMs are generally not reliable processors of raw numbers.  Tokenizers split digits in ways that do not reflect numerical magnitude, so that 1000 and 1001 may receive very different internal representations despite being nearly identical numerically.  Moreover, round numbers such as 1000 appear far more frequently in training corpora and therefore carry richer associations than their neighbours, introducing an undesirable asymmetry.  The variable names themselves (``DrivAge'', ``VehPower'') are abbreviations that may not align well with the LLM's vocabulary, and the absence of descriptive context around each value means that the model must rely entirely on pre-trained knowledge to interpret the numbers.  We nonetheless include this format because its minimal token footprint makes it the most computationally efficient option.

\subsection{Task-instruction list}\label{app:prompt-instruction}

\begin{figure}[ht]
  \centering
  \begin{prompt}
Represent this motor insurance policyholder for claim
frequency prediction:
- Age: 18 years
- Area: Urban area (region Centre)
- Population density: 1020 people/km2
- Vehicle: Renault, Nissan, or Citroën, 1 year old
- Fuel type: Diesel
- Power class: 7
- Bonus-Malus score: 100
\end{prompt}
\caption{Task-instruction list prompt.}
\label{fig:prompt-instruction}
\end{figure}

This format (Figure~\ref{fig:prompt-instruction}) prepends a short task description to the simple structured list.  It is distinct from the default underwriting-list prompt in Figure~\ref{fig:prompt-baseline}, which uses a fuller underwriting instruction and additional numeric anchors.  Instruction-aware embedding models such as those described in \citep{su2023instructor} and Qwen3-Embedding \citep{zhang2025qwen3} accept a task instruction that guides the model to produce embeddings tailored to a specific downstream use; the instruction ``Represent this motor insurance policyholder for claim frequency prediction'' signals that the embedding should emphasize risk-relevant features rather than generic semantic content.

\subsection{Qualitative descriptors}\label{app:prompt-qualitative}

\begin{figure}[ht]
  \centering
  \begin{prompt}
- Driver age: very young, newly licensed (18 years)
- Area: high-density urban (Centre region)
- Population density: high-density urban, well above the national average (1020 people/km2)
- Vehicle: Renault, Nissan, or Citroën, nearly new (1 year old)
- Fuel type: Diesel
- Vehicle power: moderate (class 7 out of 15)
- Bonus-Malus: entrance level, no prior claims (score 100)
\end{prompt}
\caption{Qualitative descriptor prompt.}
\label{fig:prompt-qualitative}
\end{figure}

This format (Figure~\ref{fig:prompt-qualitative}) replaces raw numeric values with contextual qualitative descriptions while retaining the original values in parentheses.  The motivation is that LLMs tokenize numbers in unpredictable ways and lack the ability to reason reliably about numeric magnitudes.  A population density of 1,020 people/km$^2$ is tokenized as arbitrary digit sequences (``1'', ``020'' or ``10'', ``20'', depending on the tokenizer), and the LLM has no built-in sense of whether this value is high or low relative to the portfolio distribution.  By contrast, the phrase ``high-density urban'' supplies an explicit qualitative scale, and a coarse comparison phrase (here ``well above the national average,'') places the value relative to a reference level; whether this representation helps is tested in Section~\ref{sec:prompt-format}.  Similarly, a BonusMalus score of 100 is meaningless without knowing that the scale runs from 50 (best) to 230 (worst) with 100 as the entrance level.  The descriptor ``entrance level, no prior claims'' communicates this context directly.  The qualitative binning of continuous variables (e.g., ``very young'' for drivers under 25, ``moderate'' for power classes 5--9) uses fixed, hand-set thresholds chosen from actuarial domain knowledge; they are not estimated from any data split, so they introduce no leakage into the small-$n$ experiments.  Table~\ref{tab:qualitative-bins} reports the textual classifications.

\begin{table}[ht]
  \centering
  \caption{Qualitative-descriptor bin edges used in the qualitative-descriptor prompt.}
  \label{tab:qualitative-bins}
  \setlength{\tabcolsep}{4pt}
  \begin{tabularx}{\textwidth}{@{}>{\raggedright\arraybackslash}p{1.7in}X@{}}
    \toprule
    Field & Descriptor bins (by raw value) \\
    \midrule
    Driver age & $<25$ very young, newly licensed; $<31$ young; $\leq 50$ experienced adult; $\leq 70$ older experienced; otherwise senior \\
    Population density (level) & $<100$ low; $<500$ moderate; $<1{,}000$ elevated; otherwise high \\
    Density vs.\ national average & $<100$ well below; $<500$ below; $<1{,}000$ near; otherwise well above the national average \\
    Vehicle age & $\leq 1$ nearly new; $\leq 5$ recent; $\leq 12$ mature; otherwise old \\
    Vehicle power & $\leq 4$ low; $\leq 9$ moderate; otherwise high \\
    Bonus-Malus & $<80$ strong bonus; $<100$ bonus; $=100$ entrance level; $\leq 150$ malus; otherwise severe malus \\
    \bottomrule
  \end{tabularx}
\end{table}

This format requires a preprocessing step that maps raw values to qualitative categories, but this mapping is deterministic and interpretable, and it may be viewed as a form of domain-informed feature engineering applied at the prompt level rather than at the covariate level.

\subsection{Underwriting context}\label{app:prompt-underwriting}

\begin{figure}[ht]
  \centering
  \begin{prompt}
    Represent this motor insurance policyholder for claim frequency prediction.

    Use general underwriting intuition when forming the representation. Risk is usually higher for very young drivers, high-powered or sporty vehicles, poor Bonus-Malus scores, and dense urban environments with more traffic interactions. Dense areas may raise claim frequency through congestion, intersections, and minor collisions, while lower-density areas may involve fewer crashes but potentially higher speeds. Risk is usually lower for mature drivers, modest low-powered vehicles, favourable Bonus-Malus scores, and stable driving histories. For example, a very young driver with a powerful vehicle in a dense urban area and a high Bonus-Malus score should be represented as risky, whereas a mature driver with a low-powered vehicle and a favourable Bonus-Malus score should be represented as safer.

    Now represent the following specific policyholder based only on the listed information:

    - Policyholder age: 18 years old
    - Area type: Urban area
    - Region: Centre, France
    - Population density: 1020 people/km2
    - Vehicle brand group: Renault, Nissan, or Citroën
    - Vehicle age: 1 year old
    - Fuel type: Diesel
    - Vehicle power class: 7 out of 15
    - Bonus-Malus score: 100, where 100 is the entrance level, values below 100 indicate a bonus, and values above 100 indicate a malus


\end{prompt}
\caption{Long underwriting prompt.}
\label{fig:prompt-underwriting-context}
\end{figure}

The long underwriting prompt (Figure~\ref{fig:prompt-underwriting-context}) tests a different intervention from the qualitative-descriptor template.  It provides a general actuarial frame before the policyholder fields are listed, while the qualitative-descriptor template translates each numeric value into a local descriptor.  Its purpose is to determine whether broad underwriting knowledge, stated explicitly in the input text, helps the embedding model organize otherwise unchanged covariates along risk-relevant directions.  

\section{Covariate mappings}\label{app:mappings}

The categorical covariates in the French MTPL dataset use anonymized codes. From a modelling perspective, this does not matter for GLMs, since these categories are converted into one-hot vectors. However, the categorical information does matter for LLMs, so the prompts should include the textual information corresponding to each category.  Table~\ref{tab:brand-mapping} reports the vehicle brand mapping, Table~\ref{tab:region-mapping} the region mapping, and Table~\ref{tab:area-mapping} the area mapping.  These mappings are used in the serialization step to replace raw codes with human-readable descriptions in the policyholder prompts.

\begin{table}[ht]
  \centering
  \caption{Vehicle brand code mapping.}
  \label{tab:brand-mapping}
  \begin{tabular}{ll@{\hskip 2em}ll}
    \toprule
    Code & Description & Code & Description \\
    \midrule
    B1  & Renault, Nissan, or Citro\"en & B6  & Fiat \\
    B2  & Renault, Nissan, or Citro\"en & B10 & Mercedes, Chrysler, or BMW \\
    B3  & Volkswagen, Audi, Skoda, or Seat & B11 & Mercedes, Chrysler, or BMW \\
    B4  & Opel, General Motors, or Ford & B12 & Japanese (except Nissan) or Korean \\
    B5  & Opel, General Motors, or Ford & B13 & Other \\
    &                               & B14 & Other \\
    \bottomrule
  \end{tabular}
\end{table}

\begin{table}[ht]
  \centering
  \caption{Region code mapping to French administrative regions (pre-2016 boundaries).}
  \label{tab:region-mapping}
  \begin{tabular}{ll@{\hskip 2em}ll}
    \toprule
    Code & Region & Code & Region \\
    \midrule
    R11 & \^Ile-de-France          & R52 & Pays de la Loire \\
    R21 & Champagne-Ardenne        & R53 & Bretagne \\
    R22 & Picardie                 & R54 & Poitou-Charentes \\
    R23 & Haute-Normandie          & R72 & Aquitaine \\
    R24 & Centre                   & R73 & Midi-Pyr\'en\'ees \\
    R25 & Basse-Normandie          & R74 & Limousin \\
    R26 & Bourgogne                & R82 & Rh\^one-Alpes \\
    R31 & Nord-Pas-de-Calais       & R83 & Auvergne \\
    R41 & Lorraine                 & R91 & Languedoc-Roussillon \\
    R42 & Alsace                   & R93 & Provence-Alpes-C\^ote d'Azur \\
    R43 & Franche-Comt\'e          & R94 & Corse \\
    \bottomrule
  \end{tabular}
\end{table}

\begin{table}[ht]
  \centering
  \caption{Area code mapping.}
  \label{tab:area-mapping}
  \begin{tabular}{ll}
    \toprule
    Code & Description \\
    \midrule
    A & Rural area \\
    B & Semi-rural area \\
    C & Suburban-fringe area \\
    D & Suburban area \\
    E & Urban area \\
    F & Urban center \\
    \bottomrule
  \end{tabular}
\end{table}

\end{document}